\documentclass[fleqn]{article}
\usepackage{longtable}
\usepackage{graphicx}

\begin {document}
\begin{flushleft}
{\LARGE
{\bf Energy levels, radiative rates and electron impact excitation rates for transitions in Be-like Ti XIX}
}\\

\vspace{1.5 cm}

{\bf {Kanti  M  ~Aggarwal and Francis  P  ~Keenan}}\\ 

\vspace*{1.0cm}

Astrophysics Research Centre, School of Mathematics and Physics, Queen's University Belfast, Belfast BT7 1NN, Northern Ireland, UK\\ 

\vspace*{0.5 cm} 

e-mail: K.Aggarwal@qub.ac.uk \\

\vspace*{1.50cm}

Received  23 May 2012\\
Accepted for publication 12 September  2012 \\
Published xx  Month 2012 \\
Online at stacks.iop.org/PhysScr/vol/number \\

\vspace*{1.5cm}

PACS Ref: 32.70 Cs, 34.80 Dp, 95.30 Ky

\vspace*{1.0 cm}

\hrule

\vspace{0.5 cm}
{\Large {\bf S}} This article has associated online supplementary data files \\
Tables 2 and 5 are available only in the electronic version at stacks.iop.org/PhysScr/vol/number

\end{flushleft}

\clearpage


\begin{abstract}

We report calculations of energy levels, radiative rates and electron impact excitation cross sections and rates for transitions in Be-like Ti XIX.  The {\sc grasp} (General-purpose Relativistic Atomic Structure Package) is adopted for calculating energy levels and radiative rates. For determining the collision strengths and subsequently the excitation rates, the Dirac Atomic R-matrix Code ({\sc darc}) is used.  Oscillator strengths, radiative rates and line strengths are reported for all E1, E2, M1 and M2 transitions among the lowest 98 levels of the $n \le$ 4 configurations. Additionally, theoretical lifetimes are listed for all  98 levels.  Collision strengths are averaged over a Maxwellian velocity distribution and the effective collision strengths  obtained listed over a wide temperature range up to 10$^{7.7}$ K. Comparisons are made with similar data obtained from the Flexible Atomic Code ({\sc fac}) to  highlight the importance of resonances, included in calculations with {\sc darc}, in the determination of effective collision strengths. Discrepancies between the collision strengths from {\sc darc} and {\sc fac}, particularly for forbidden transitions, are also discussed.  

\end{abstract}

\clearpage

\section{Introduction}

Emission lines of Ti ions, including Ti XIX,  have been widely measured in laboratory plasmas \cite{gold}--\cite{yz}, due to their interest for the development of x-ray lasers. Titanium is also often a material in the walls of  fusion reactors, and hence  many ionisation stages of this element are observed in fusion spectra  due to the high temperatures.  Considering its importance, several calculations have been performed in the past \cite{akb1}--\cite{azza} to determine atomic data for energy levels, radiative rates (A- values), and excitation rates or equivalently the effective collision strengths ($\Upsilon$), which are obtained from the electron impact collision strengths ($\Omega$). Additionally, O'Mahony  {\em et al} \cite{mkc} have reported analytical expressions to derive values of $\Upsilon$ for Ti XIX,  based on  $R$-matrix calculations for Be-like ions between Sc XVIII and Zn XXVII. However, all these data are for transitions among the lowest 10 levels of the $n$=2 configurations of Ti XIX, and no calculation has so far been performed with the $R$-matrix code which explicitly includes the contribution of resonances in the determination of  $\Upsilon$. The resonance contribution to  $\Upsilon$ may be highly significant, particularly for the forbidden transitions, as we will demonstrate in section 6.  Therefore, in this work we report atomic data for energy levels, A- values, $\Omega$ and $\Upsilon$ for transitions among the lowest 98 levels of the $n \le$ 4 configurations of Ti XIX. 

For calculations of energy levels and A-values we employ the fully relativistic {\sc grasp} (general-purpose relativistic atomic structure  package) code,  which was originally developed by Grant {\em et al} \cite{grasp0} and revised by Dr P H Norrington. It is a fully relativistic code, and is based on the $jj$ coupling scheme.  Further relativistic corrections arising from the Breit (magnetic) interaction and quantum electrodynamics (QED) effects (vacuum polarization and Lamb shift) have also been included. Additionally, we have used the option of {\em extended average level} (EAL),  in which a weighted (proportional to 2$j$+1) trace of the Hamiltonian matrix is minimized. This produces a compromise set of orbitals describing closely lying states with  moderate accuracy. For our calculations of $\Omega$, we have adopted the {\em Dirac Atomic $R$-matrix Code} ({\sc darc}) of P H Norrington and I P Grant ({\tt http://web.am.qub.ac.uk/DARC/}). Finally, for comparison purposes, we have performed parallel calculations with the {\em Flexible Atomic Code} ({\sc fac}) of Gu \cite{fac}, available from the website {\tt {\verb+http://sprg.ssl.berkeley.edu/~mfgu/fac/+}}. This is also a fully relativistic code which provides a variety of atomic parameters, and (generally) yields results for energy levels and radiative rates comparable to {\sc grasp}  -- see, for example, Aggarwal {\em et al} \cite{fe15}.  However, differences in collision strengths and subsequently in effective collision strengths with those obtained from {\sc darc} can be large, particularly for  forbidden transitions, as demonstrated in some of our earlier papers \cite{nax}--\cite{tixxi}, and also discussed below in sections 5 and 6. Hence results from {\sc fac} will be helpful in assessing the accuracy of our energy levels and radiative rates, and in estimating the contribution of resonances to the determination of effective collision strengths, included in calculations from {\sc darc} but not in {\sc fac}.

\section{Energy levels}

The 17 configurations of Ti XIX, namely (1s$^2$) 2$\ell$2$\ell'$, 2$\ell$3$\ell'$ and 2$\ell$4$\ell'$,  give rise to the lowest 98 levels listed in Table 1, where we also provide  our level energies calculated from {\sc grasp},  {\em without} and {\em with} the inclusion of Breit and QED effects.  Wiese and Fuhr \cite{nist1} have compiled and critically evaluated experimentally measured  energy levels of Ti XIX,  listed  at the  NIST (National Institute of Standards and Technology)   website {\tt http://www.nist.gov/pml/data/asd.cfm}. These compilations are  included in Table 1 for comparisons.   However, NIST energies are not available for many levels, particularly  of the 2$\ell$4$\ell'$ configurations, and for some of the levels their results are indistinguishable  -- see for example: 26/27 [(2p3p) $^3$D$_2$ and $^1$P$_1$] and 41/42 [2p3d $^3$P$^o_{2,1}$].  Also included in the table are our calculations obtained from the {\sc fac} code (FAC1), including the same CI (configuration interaction) as in {\sc grasp}.  

Our level energies obtained without the Breit and QED effects (GRASP1) are higher than the NIST values by  up to $\sim$ 0.15 Ryd for some of the levels, such as: 9 (2p$^2$ $^1$D$_2$), 10 (2p$^2$ $^1$S$_0$) and 17 (2s3d $^3$D$_1$). Furthermore, the ordering is also mostly the same as  that of NIST.  However, there are also striking differences, in both ordering and magnitude,  for some of the levels, namely 45/46 [(2p3d) $^1$F$^o_3$ and $^1$P$^o_1$], 53/54 [2s4d $^3$D$_{1,2}$] and 85 [2d4d $^3$P$^o_2$] , for which the discrepancy is up to 0.4 Ryd.  The inclusion of Breit and QED effects (GRASP2) lowers the energies by a maximum of  $\sim$ 0.065 Ryd, indicating that for this ion the higher relativistic effects are not too important. In addition, the ordering has slightly altered in a few instances, see for example levels 37/38 [2p3p $^1$D$_2$ and 2p3d $^3$F$^o_4$] and 60/61 [2s4f $^1$F$^o_3$ and 2p4s $^3$P$^o_0$]. However, the energy differences for these swapped levels are very small. Our FAC1 level energies  agree with our GRASP2 calculations within 0.04 Ryd for all levels and the orderings are also the same.  Small differences in the {\sc grasp} and {\sc fac} energies arise  mostly by the ways calculations of central potential for radial orbitals and recoupling schemes of angular parts have been performed -- see detailed discussion in the {\sc fac} manual. A further inclusion of the 2$\ell$5$\ell'$ configurations, labelled FAC2 calculations in Table 1, makes no appreciable difference either in the magnitude or ordering of the levels. Therefore, we are confident of our energy levels listed in Table 1, and assess these to be accurate to better than 0.5\%.

\section{Radiative rates}

Since currently available  A- values in the literature are limited to transitions among the lowest 10 levels of Ti XIX, we here provide a complete set of data for all transitions among the 98 levels and for four types, namely electric dipole (E1), electric quadrupole (E2), magnetic dipole (M1), and  magnetic quadrupole (M2), as these are required in  a  plasma model. Furthermore, the absorption oscillator strength ($f_{ij}$) and radiative rate A$_{ji}$ (in s$^{-1}$) for a transition $i \to j$ are related by the following expression \cite{rhg}:

\begin{equation}
f_{ij} = \frac{mc}{8{\pi}^2{e^2}}{\lambda^2_{ji}} \frac{{\omega}_j}{{\omega}_i} A_{ji}
 = 1.49 \times 10^{-16} \lambda^2_{ji} (\omega_j/\omega_i) A_{ji}
\end{equation}
where $m$ and $e$ are the electron mass and charge, respectively, $c$ is the velocity of light,  $\lambda_{ji}$ is the transition energy/wavelength in $\rm \AA$, and $\omega_i$
and $\omega_j$ are the statistical weights of the lower ($i$) and upper ($j$) levels, respectively. Similarly, the oscillator strength f$_{ij}$ (dimensionless) and the line
strength S (in atomic unit, 1 a.u. = 6.460$\times$10$^{-36}$ cm$^2$ esu$^2$) are related by the  standard equations given below  \cite{rhg}-\cite{uis}.

\begin{flushleft}
For the electric dipole (E1) transitions 
\end{flushleft} 
\begin{equation}
A_{ji} = \frac{2.0261\times{10^{18}}}{{{\omega}_j}\lambda^3_{ji}} S^{{\rm E1}} \hspace*{0.5 cm} {\rm and} \hspace*{0.5 cm} 
f_{ij} = \frac{303.75}{\lambda_{ji}\omega_i} S^{{\rm E1}}, \\
\end{equation}
\begin{flushleft}
for the magnetic dipole (M1) transitions  
\end{flushleft}
\begin{equation}
A_{ji} = \frac{2.6974\times{10^{13}}}{{{\omega}_j}\lambda^3_{ji}} S^{{\rm M1}} \hspace*{0.5 cm} {\rm and} \hspace*{0.5 cm}
f_{ij} = \frac{4.044\times{10^{-3}}}{\lambda_{ji}\omega_i} S^{{\rm M1}}, \\
\end{equation}
\begin{flushleft}
for the electric quadrupole (E2) transitions 
\end{flushleft}
\begin{equation}
A_{ji} = \frac{1.1199\times{10^{18}}}{{{\omega}_j}\lambda^5_{ji}} S^{{\rm E2}} \hspace*{0.5 cm} {\rm and} \hspace*{0.5 cm}
f_{ij} = \frac{167.89}{\lambda^3_{ji}\omega_i} S^{{\rm E2}}, 
\end{equation}

\begin{flushleft}
and for the magnetic quadrupole (M2) transitions 
\end{flushleft}
\begin{equation}
A_{ji} = \frac{1.4910\times{10^{13}}}{{{\omega}_j}\lambda^5_{ji}} S^{{\rm M2}} \hspace*{0.5 cm} {\rm and} \hspace*{0.5 cm}
f_{ij} = \frac{2.236\times{10^{-3}}}{\lambda^3_{ji}\omega_i} S^{{\rm M2}}. \\
\end{equation}

In Table 2 we present transition energies/wavelengths ($\lambda$, in $\rm \AA$), radiative rates (A$_{ji}$, in s$^{-1}$), oscillator strengths (f$_{ij}$, dimensionless), and line strengths (S, in a.u.), in length  form only, for all 1468 electric dipole (E1) transitions among the 98 levels of Ti XIX. The {\em indices} used  to represent the lower and upper levels of a transition have already been defined in Table 1. Similarly, there are 1754 electric quadrupole (E2), 1424  magnetic dipole (M1), and 1792 magnetic quadrupole (M2) transitions among the 98 levels. However, for these transitions only the A-values are listed in Table 2, and the corresponding results for f- or S- values can be easily obtained using Eqs. (1--5).

As noted earlier,  A-values in the literature for Ti XIX  are only available  for a limited number of  transitions.  Therefore, we have performed another calculation with the {\sc fac} code of Gu \cite{fac}.  In Table 3 we compare our A- values from both the {\sc grasp} and {\sc fac} codes for some transitions among the lowest 20 levels of Ti XIX. Also included in this table are f- values from {\sc grasp} because they give an indication of the strength of a transition. Similarly, to facilitate easy comparison between the two calculations, we have also listed the ratio of A-values obtained with the {\sc grasp} and {\sc fac} codes. For these (and many other) transitions, the agreement between the two sets of A- values is better than 20\%. Indeed, for  most  strong transitions (f $\ge$ 0.01), the A- values from {\sc grasp} and {\sc fac} agree to better than 20\%, and the only exceptions are three transitions, namely 2--32 (2s2p $^3$P$^o_0$ -- 2p3p $^3$S$_1$), 32--71 (2p3p $^3$S$_1$ -- 2p4d $^3$D$^o_2$) and 32--83 (2p3p $^3$S$_1$ -- 2p4d $^1$D$^o_2$), for which the discrepancies are up to 40\%.  These discrepancies mainly arise from the corresponding differences in the energy levels. Furthermore, for a majority (80\%) of the strong E1 transitions (f $\ge$ 0.01) the length and velocity forms in our {\sc grasp} calculations agree within 20\%, and discrepancies for the others are mostly within a factor of two.  However, for a few ($\sim$ 13\%) weaker transitions (f $\le$ 10$^{-3}$) the two forms of the f- value differ by up to several  orders of magnitude, and examples include: 4--24 (f $\sim$ 3$\times$10$^{-10}$), 4--92 (f $\sim$ 4$\times$10$^{-7}$), 29--31 (f $\sim$ 5$\times$10$^{-9}$), 30--31 (f $\sim$ 6$\times$10$^{-7}$) and 33--34 (f $\sim$ 3$\times$10$^{-6}$). Finally, as for the energy levels, the effect of additional CI is negligible on the A- values, as results for all strong E1 transitions agree within $\sim$ 20\% with those obtained with the  inclusion of the $n$ = 5 configurations. To conclude,  we may state that for almost all strong E1 transitions, our radiative rates are accurate to better than 20\%. However, for the weaker transitions the accuracy is comparatively poorer.

\section{Lifetimes}

The lifetime $\tau$ for a level $j$ is defined as follows \cite{wood}:

\begin{equation}  {\tau}_j = \frac{1}{{\sum_{i}^{}} A_{ji}}.  
\end{equation} 
 Since this is a measurable parameter, it provides a check on the accuracy of the calculations. Therefore, in Table 1 we have also listed our calculated lifetimes, which include the contributions from four types of transitions, i.e. E1, E2, M1, and M2. To our knowledge, no calculations or measurements are available for lifetimes for any of the Ti XIX levels. However, we hope the present results will be useful for future comparisons and may encourage experimentalists to measure lifetimes, particularly for the level 2s2p $^3$P$^o_2$  which has a  comparatively large value of $\sim$ 1 ms.

\section{Collision strengths}

Collision strengths ($\Omega$) are related to the more commonly known parameter collision cross section ($\sigma_{ij}$, $\pi{a_0}^2$) by the following relationship \cite{bt}:

\begin{equation}
\Omega_{ij}(E) = {k^2_i}\omega_i\sigma_{ij}(E)
\end{equation}
where ${k^2_i}$ is the incident energy of the electron and $\omega_i$ is the statistical weight of the initial state. Results for collisional data are preferred 
in the form of $\Omega$ because it is a symmetric and dimensionless quantity.

For the computation of collision strengths $\Omega$, we have employed the {\em Dirac atomic $R$-matrix code} ({\sc darc}), which includes the relativistic effects in a
systematic way, in both the target description and the scattering model. It is based on the $jj$ coupling scheme, and uses the  Dirac-Coulomb Hamiltonian in the $R$-matrix
approach. The $R$-matrix radius adopted for Ti XIX is 3.64 au, and 55  continuum orbitals have been included for each channel angular momentum in the expansion of the wavefunction, allowing us to compute $\Omega$ up to an energy of  1150 Ryd, i.e. $\sim$ 1070 Ryd {\em above} the highest threshold,  equivalent to $\sim$ 1.7$\times$10$^8$ K. This energy range is  sufficient to calculate values of effective collision strength $\Upsilon$ (see section 6)  up to T$_e$ = 10$^{7.7}$ K, well above the temperature of maximum abundance in ionisation equilibrium for Ti XIX, i.e. 10$^{6.9}$ K  \cite{pb}.  The maximum number of channels for a partial wave is 428, and the corresponding size of the Hamiltonian matrix is 23 579. To obtain convergence of  $\Omega$ for all transitions and at all energies, we have included all partial waves with angular momentum $J \le$ 40.5, although a larger number would have been  preferable for the convergence of some allowed transitions, especially at higher energies. However, to account for higher neglected partial waves, we have included a top-up, based on the Coulomb-Bethe approximation \cite{ab} for allowed transitions and geometric series for others.

For illustration, in Figs. 1-3 we show the variation of $\Omega$ with angular momentum $J$ for three transitions of Ti XIX, namely 1--5 (2s$^2$ $^1$S$_0$ -- 2s2p $^1$P$^o_1$), 2--4 (2s2p $^3$P$^o_0$ -- 2s2p $^3$P$^o_2$)  and 9--10 (2p$^2$  $^1$D$_0$ -- 2p$^2$ $^1$S$_0$),  and at three energies of 100, 500 and 900 Ryd. The values of $\Omega$ have not converged for allowed transitions  as shown in Fig. 1,  for which a top-up has been included as mentioned above, and has been found to be appreciable.  However, for all forbidden transitions,  the values of $\Omega$ have fully converged as shown in Figs. 2 and 3. It is also clear from Figs. 2 and 3 that a large range of partial waves is required for the convergence of $\Omega$ for some of the forbidden transitions, particularly towards the higher end of the energy range.

In Table 4 we list our values of $\Omega$ for resonance transitions of Ti XIX at energies {\em above} thresholds. The  indices used  to represent the levels of a transition have already been defined in Table 1. Unfortunately, no similar data are available for comparison purposes as already noted in section 1. Therefore,  to make an accuracy assessment for $\Omega$, we have performed another calculation using the {\sc fac} code of Gu \cite{fac}. This code is also fully relativistic, and is based on the well-known and widely-used {\em distorted-wave} (DW) method -- see for example, \cite{dw1}--\cite{dw2} and the {\sc FAC} manual.  Furthermore, the same CI is included in {\sc fac} as in the calculations from {\sc darc}. Therefore, also included in Table 4 for comparison purposes are the $\Omega$ values from {\sc fac} at a single {\em excited} energy E$_j$, which corresponds to an incident energy of $\sim$ 700 Ryd for Ti XIX.  For $\sim$ 60\% of the Ti XIX transitions, the values of $\Omega$ with the {\sc darc} and {\sc fac} codes agree within 20\% at an energy of 700 Ryd. However,  the discrepancies for others are much higher, particularly for weaker transitions, such as:  1--30/31/35/39/73/77/83/89/91. Most of these are weak ($\Omega \le$ 10$^{-6}$) and forbidden, i.e. the values of $\Omega$  have fully converged at {\em all} energies within our adopted range of partial waves in the calculations with the {\sc darc} code. For such weak transitions, values of $\Omega$ from the {\sc fac} code are not assessed to be accurate.  Additionally, for a few transitions, such as 49--87, 50--72/86/98, 51--66/69/71/83/85, 52--68/72/86/98 and 53--63/64/77/95,  the values of $\Omega$ from the {\sc fac} code show a sudden increase,  by orders of magnitude  at some random energies, generally towards the higher end. This problem is common for many ions and  examples of this can be seen in Fig. 6 of Aggarwal and Keenan \cite{mgxi}, \cite{caxix}. The sudden anomalous behaviour in  $\Omega$ with the {\sc fac} code is also responsible for the differences noted above for many  transitions. Such anomalies for some transitions (both allowed and forbidden)  from the  {\sc fac} calculations  arise primarily because of the interpolation and extrapolation techniques employed in the  code. In order to expedite calculations, i.e.  to generate a large amount of atomic data in a comparatively very short period of time, and without too large loss of accuracy, calculations of $\Omega$ are not performed at each partial wave, but only at each $J$ up to 5, and then the interval between successive calculations is doubled every two points, i.e. the grid is almost logarithmic -- see the {\sc fac} manual for further details. Similarly, some differences in  $\Omega$ are expected because the DW method generally overestimates  results due to the exclusion of channel coupling.

As a further comparison between the {\sc darc} and {\sc fac} values of $\Omega$, in Fig. 4 we show the variation of $\Omega$ with energy for three {\em allowed} transitions among the excited levels of Ti XIX, namely 4--19 ( 2s2p $^3$P$^o_2$ -- 2s3d $^3$D$_3$), 5--20 (2s2p $^1$P$^o_1$ -- 2s3d $^1$D$_2$), and 8--40 (2p$^2$ $^3$P$_2$ -- 2p3d $^3$D$^o_3$). For  many  transitions there are no discrepancies between the f- values obtained with the two different codes ({\sc grasp} and {\sc fac}) as demonstrated in Table 3, and therefore the values of $\Omega$ also agree to better than 20\%.  Similar comparisons between the two calculations with {\sc darc} and {\sc fac} are shown in Fig. 5 for three {\em forbidden} transitions of Ti XIX, namely 1--12 (2s$^2$ $^1$S$_0$ -- 2s3s $^1$S$_0$), 2--4 (2s2p $^3$P$^o_0$ -- 2s2p $^3$P$^o_2$), and 3--4 (2s2p $^3$P$^o_1$ -- 2s2p $^3$P$^o_2$). As in the case of  the allowed transitions, for these forbidden ones  the agreement between the two calculations is generally satisfactory, although there are some  differences  towards the lower end of the energy range.  Therefore, on the basis of these and other comparisons discussed above, collision strengths from our {\sc darc} code are  assessed to be accurate to better than 20\%.  However, similar data from {\sc fac}  are not assessed to be accurate for all transitions over an entire energy range.

\section{Effective collision strengths}

Excitation rates, in addition to energy levels and radiative rates, are required for plasma modelling, and are determined from the collision strengths ($\Omega$). Since the
threshold energy region is dominated by numerous closed-channel (Feshbach) resonances, values of $\Omega$ need to be calculated in a fine energy mesh  to accurately
account for their contribution. Furthermore, in a plasma electrons have a wide distribution of velocities, and therefore values of $\Omega$ are generally averaged over a
{\em Maxwellian} distribution as follows  \cite{bt}:

\begin{equation}
\Upsilon(T_e) = \int_{0}^{\infty} {\Omega}(E) {\rm exp}(-E_j/kT_e) d(E_j/{kT_e}),
\end{equation}
where $k$ is Boltzmann constant, T$_e$  electron temperature in K, and E$_j$  the electron energy with respect to the final (excited) state. Once the value of $\Upsilon$ is
known the corresponding results for the excitation q(i,j) and de-excitation q(j,i) rates can be easily obtained from the following equations:

\begin{equation}
q(i,j) = \frac{8.63 \times 10^{-6}}{{\omega_i}{T_e^{1/2}}} \Upsilon {\rm exp}(-E_{ij}/{kT_e}) \hspace*{1.0 cm}{\rm cm^3s^{-1}}
\end{equation}
and
\begin{equation}
q(j,i) = \frac{8.63 \times 10^{-6}}{{\omega_j}{T_e^{1/2}}} \Upsilon \hspace*{1.0 cm}{\rm cm^3 s^{-1}},
\end{equation}
where $\omega_i$ and $\omega_j$ are the statistical weights of the initial ($i$) and final ($j$) states, respectively, and E$_{ij}$ is the transition energy. The contribution of
resonances may enhance the values of $\Upsilon$ over those of the background  collision strengths ($\Omega_B$), especially for the forbidden transitions, by up to an
order of magnitude (or even more) depending on the transition and/or the temperature.  Similarly, values of $\Omega$ need to be calculated over a wide energy range (above
thresholds)  to obtain convergence of the integral in Eq. (8), as demonstrated in Fig. 7 of Aggarwal and Keenan \cite{ni11a}. It may be noted that if for practical reasons calculations of $\Omega$ are performed only up to a limited range of energy then the high energy limits for a range of transitions can be invoked through the expressions suggested by Burgess and Tully \cite{bt}. However, there is no such need in the present work as calculations for $\Omega$ have been performed up to a reasonably high energy range, as noted in section 5. 

To delineate resonances, we have performed our calculations of $\Omega$ at over $\sim$ 41 000 energies in the thresholds region. Close to thresholds ($\sim$0.1 Ryd above a threshold) the energy mesh is 0.001 Ryd, and away from thresholds is 0.002 Ryd. Hence care has been taken to include as many resonances as possible, and with as fine a resolution as  is computationally feasible. The density and importance of resonances can be appreciated from Figs. 6--11, where we plot $\Omega$ as a function of energy in the thresholds region for the  1--2 (2s$^2$ $^1$S$_0$ -- 2s2p $^3$P$^o_0$), 1--3 (2s$^2$ $^1$S$_0$ -- 2s2p $^3$P$^o_1$), 1--4 (2s$^2$ $^1$S$_0$ -- 2s2p $^3$P$^o_2$), 2--3 (2s2p $^3$P$^o_0$ -- 2s2p $^3$P$^o_1$), 2--4 (2s2p $^3$P$^o_0$ -- 2s2p $^3$P$^o_2$) and 3--4 (2s2p $^3$P$^o_1$ -- 2s2p $^3$P$^o_2$) transitions, respectively. For all these (and many other) transitions  the resonances are dense over the entire thresholds energy range, and hence make a significant contribution to  $\Upsilon$  over a wide range of  temperatures. Since for many transitions, resonances are dense and have high magnitude at energies close to the thresholds, a slight displacement in their positions can significantly affect the calculations of $\Upsilon$,  mostly  at the low temperatures, but not at the higher ones  required for Ti XIX.

Our calculated values of $\Upsilon$ are listed in Table 5 over a wide temperature range up to 10$^{7.7}$ K, suitable for applications to a variety of plasmas. Corresponding data at any other temperature/s  and/or in a different format in a machine readable form can also be requested from any one of the authors. As stated in section 1, there are only limited results available for comparison purposes. Therefore, we have also calculated values of $\Upsilon$ from our non-resonant $\Omega$ data obtained with the {\sc fac} code, and these are included at the lowest and the highest calculated temperatures. These calculations are particularly helpful in  providing an estimate of the importance of resonances in the determination of excitation rates.  Furthermore, Zhang and Sampson \cite{zs92} (ZS) have reported values of $\Upsilon$ for transitions among the lowest 10 levels of Ti XIX. In their calculations, they have adopted the Coulomb-Born-Exchange method and their results are stored in the {\sc chianti} database at {\tt http://www.chiantidatabase.org/chianti\_direct\_data.html}. In Table 6 we compare our results for $\Upsilon$, from both {\sc darc} and {\sc fac},  with those of ZS at three  temperatures of 10$^{6.3}$, 10$^{6.9}$ and 10$^{7.5}$ K. Values of $\Upsilon$  from {\sc fac} generally agree with those of ZS within 20\%, because both calculations are based on the DW method and do not include the contribution of resonances. However, our corresponding results from {\sc darc} are higher, by up to a factor of four, for many transitions, mostly forbidden.  This is because of the inclusion of resonances in the calculations with {\sc darc}. Moreover, since resonances are spread over a wide energy range, as noted in Figs. 6--11, higher values of $\Upsilon$ are sustained over the entire range of temperatures over which the calculations have been performed -- see for example transitions 1--2/4/6/7/8/9/10.

The comparison of $\Upsilon$  in Table 6 is limited to the 45 transitions among the lowest 10 levels of Ti XIX. For a larger range of transitions, about half  have a discrepancy of more than 20\%  over the entire range of temperatures. At lower temperatures, the differences are generally within a factor of 5, but are higher (up to two orders of magnitude) for some, such as: 2--38 (2s2p $^3$P$^o_0$ -- 2p3d $^3$F$^o_4$), 6--11 (2p$^2$ $^3$P$_0$ -- 2s3s $^3$S$_1$), 7--11 (2p$^2$ $^3$P$_1$ -- 2s3s $^3$S$_1$), 8--11 (2p$^2$ $^3$P$_2$ -- 2s3s $^3$S$_1$)  and 9--51 (2p$^2$ $^1$D$_2$ -- 2s4p $^3$P$^o_2$). Similarly, towards  higher  temperatures, the discrepancies for most transitions are within a factor of two, but are larger by up to  orders of magnitude for a few, such as: 2--38/43, 6--11/14/16/17 and 7--11/18/19. In most cases, our results from {\sc darc} are higher because of the inclusion of resonances. However, in a few cases the values of $\Upsilon$  from  {\sc fac}  are abnormally greater because of the anomaly in the calculated values of $\Omega$, as discussed in section 5.

\section{Conclusions}

In this paper we have presented results for energy levels and  radiative rates for four types of transitions (E1, E2, M1, and M2) among the lowest 98 levels of Ti XIX belonging to the $n \le$ 4 configurations. Additionally, lifetimes of all the calculated levels have been reported, although  no measurements or other theoretical results are available for comparison.  However, based on a variety of comparisons among various calculations with the {\sc grasp} and {\sc fac} codes, our results for radiative rates, oscillator strengths, line strengths, lifetimes and  collision strengths are judged to be accurate to better than 20\% for a majority of the strong transitions (levels). Furthermore, for  calculations of $\Upsilon$, resonances in the thresholds energy region are noted to be dominant for many transitions, and inclusion of their contribution has significantly enhanced the results. In the absence of other similar calculations, it is difficult to fully assess the accuracy of our $\Upsilon$ results.  However, since  we have considered a large range of partial waves to achieve convergence of $\Omega$ at all energies, included a wide energy range to  calculate  values of $\Upsilon$ up to T$_e$ = 10$^{7.7}$ K, and resolved resonances in a fine energy mesh to account for their contributions, we see no apparent deficiency in our reported data.  Therefore, based on the comparisons made in section 6 and our past experience with calculations on other ions, we estimate the accuracy of our results for  $\Upsilon$  to be better than 20\% for most transitions.  Nevertheless, the present data for $\Upsilon$  for transitions involving the levels of the $n$ = 4 configurations may perhaps be improved by the inclusion of the levels of the $n$ = 5 configurations. Furthermore, for some highly charged ions, particularly He-like, the effect of radiation damping may reduce the contribution of resonances in the determination of effective collision strengths. While this may be true for a few transitions towards the lower end of the temperature range, as demonstrated by several workers, see for example:  \cite{dpz}--\cite{damp}, the effect is not appreciable at high temperatures at which data are applied in the modelling of plasmas, as discussed in some of our earlier papers, such as: \cite{mgxi}--\cite{tixxi} and \cite{kr35}.  Nevertheless, scope remains for improvement in the reported data but until then we believe the present set of complete results for radiative and excitation rates for transitions in Ti XIX will be highly useful for the modelling of a variety of plasmas. 




\clearpage

%
\begin{figure*}
 \includegraphics[scale=0.70,angle=-90]{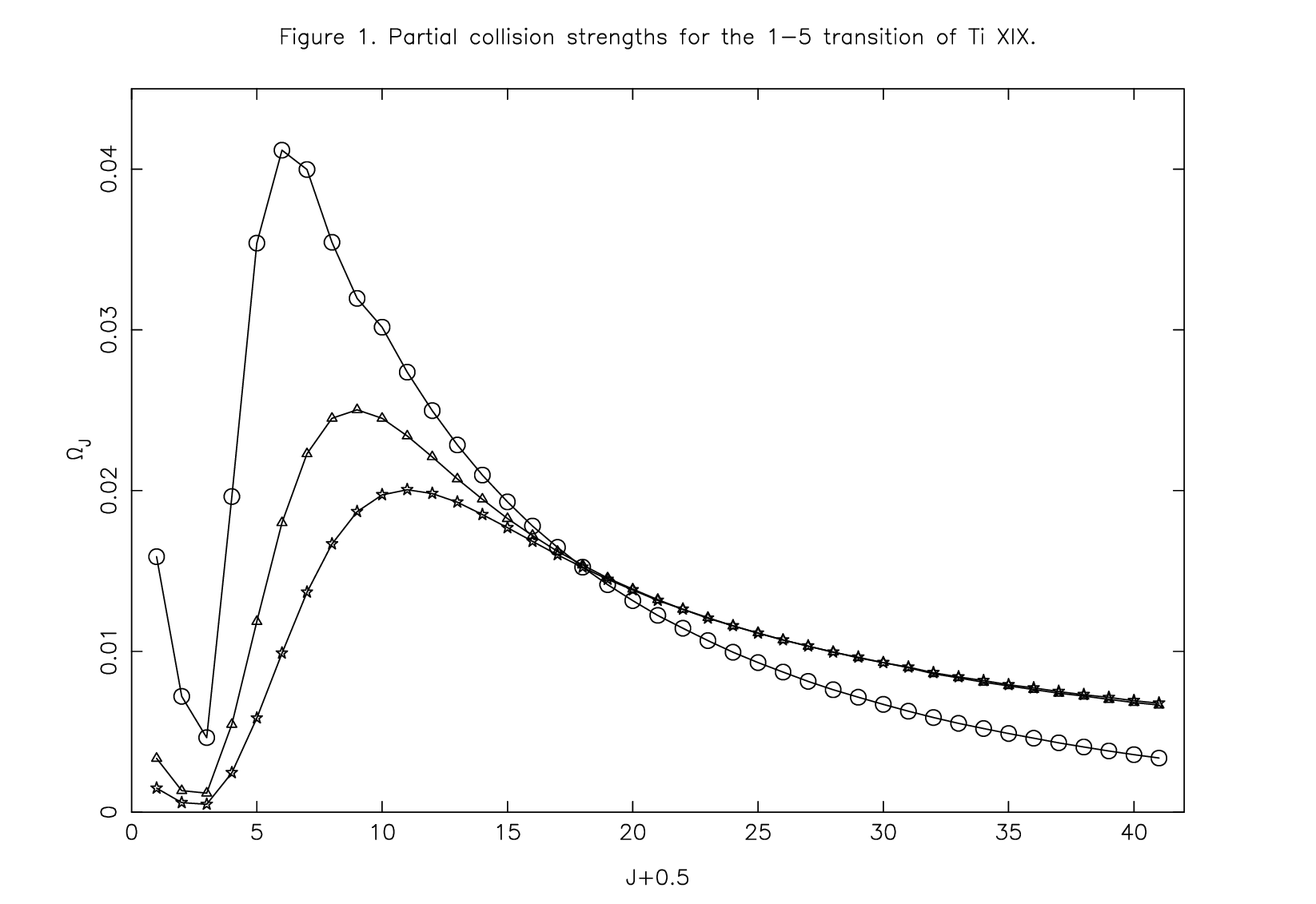}
\caption{Partial collision strengths for the 2s$^2$ $^1$S$_0$ -- 2s2p $^1$P$^o_1$  (1--5) transition of Ti XIX, 
at three energies of: 100 Ryd (circles), 500 Ryd (triangles) and 900 Ryd (stars).}
\label{fig:1}       
\end{figure*}

\begin{figure*}
\includegraphics[scale=0.70,angle=-90]{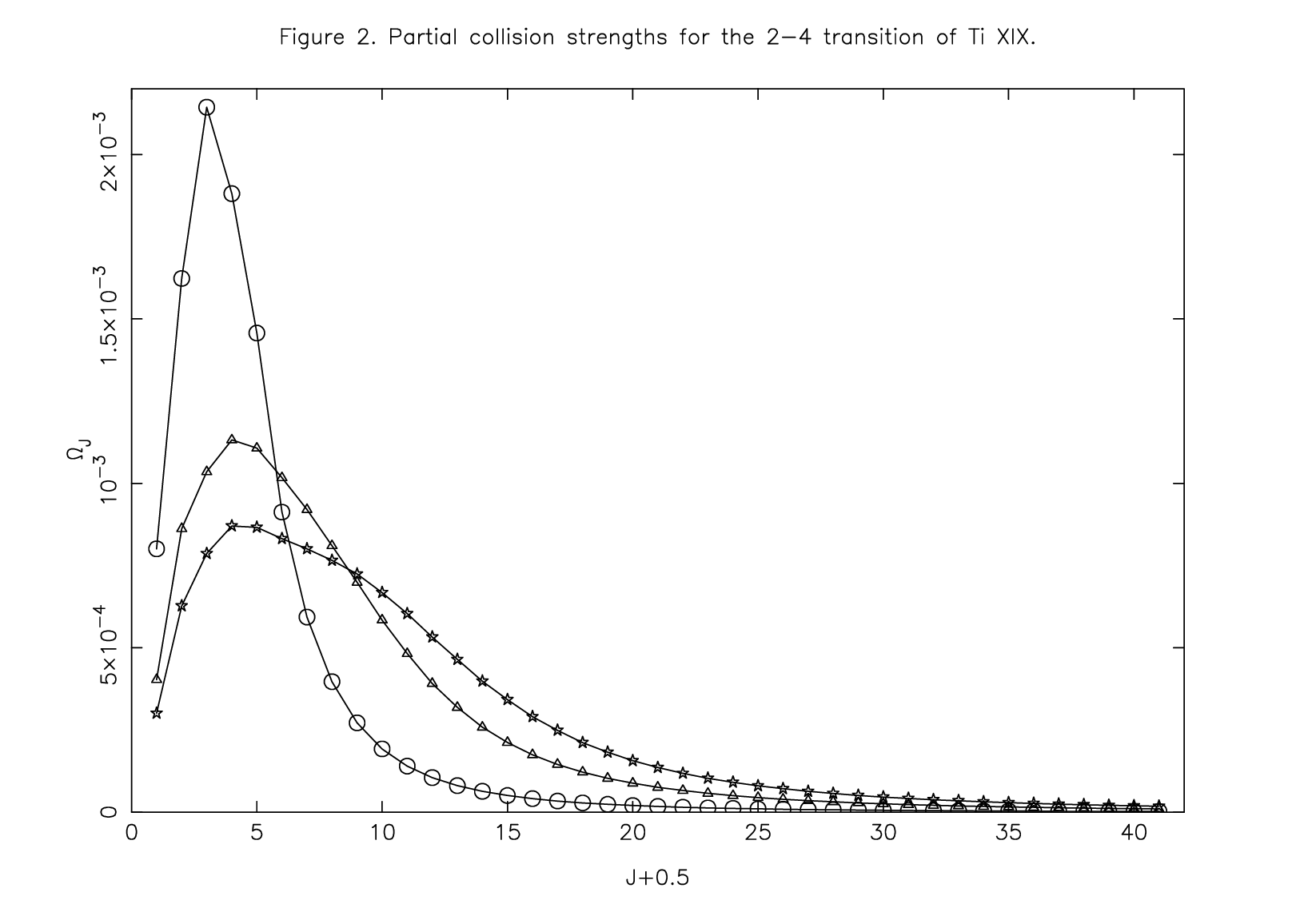}
\caption{Partial collision strengths for the 2s2p $^3$P$^o_0$ -- 2s2p $^3$P$^o_2$ (2--4)  transition of Ti XIX, 
at three energies of: 100 Ryd (circles), 500 Ryd (triangles) and 900 Ryd (stars).}
\label{fig:2}       
\end{figure*}
\clearpage
%

\begin{figure*}
\includegraphics[scale=0.70,angle=-90]{fig3.ps}
\caption{Partial collision strengths for the 2p$^2$  $^1$D$_0$ -- 2p$^2$ $^1$S$_0$ (9--10)  transition of Ti XIX, 
at three energies of: 100 Ryd (circles), 500 Ryd (triangles) and 900 Ryd (stars).}
\label{fig:3}       
\end{figure*}

\clearpage

\begin{figure*}
\includegraphics[scale=0.70,angle=-90]{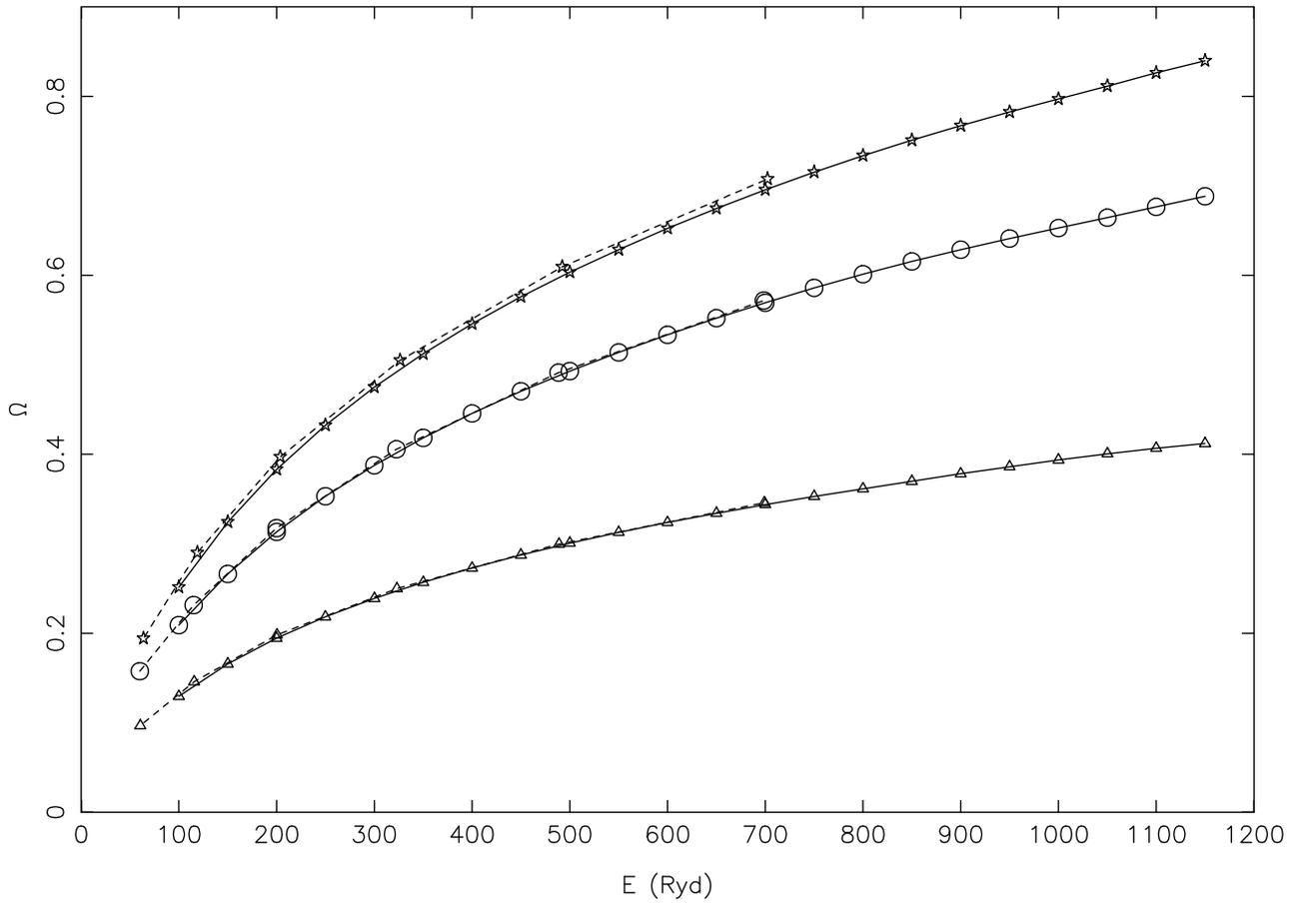}
\caption{Comparison of collision strengths from our calculations from {\sc darc} (continuous curves) and {\sc fac} (broken curves) for the  4--19 (circles: 2s2p $^3$P$^o_2$ -- 2s3d $^3$D$_3$), 5--20 (triangles: 2s2p $^1$P$^o_1$ -- 2s3d $^1$D$_2$), and 8--40 (stars: 2p$^2$ $^3$P$_2$ -- 2p3d $^3$D$^o_3$) allowed transitions of Ti XIX.}
\label{fig:4}       
\end{figure*}

\clearpage

\begin{figure*}
\includegraphics[scale=0.70,angle=-90]{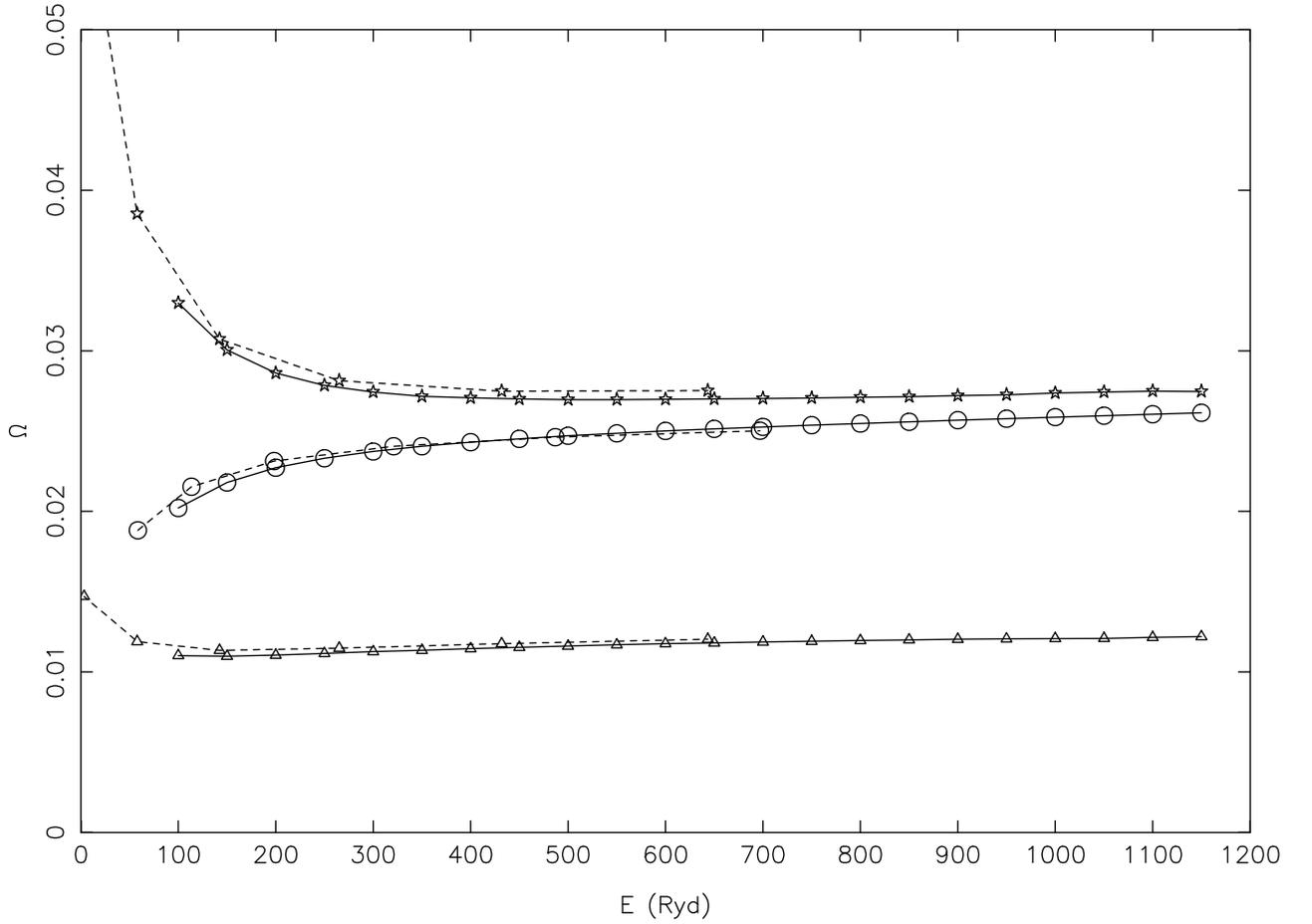}
\caption{Comparison of collision strengths from our calculations from {\sc darc} (continuous curves) and {\sc fac} (broken curves) for the  1--12 (circles: 2s$^2$ $^1$S$_0$ -- 2s3s $^1$S$_0$), 2--4 (triangles: 2s2p $^3$P$^o_0$ -- 2s2p $^3$P$^o_2$), and 3--4 (stars: 2s2p $^3$P$^o_1$ -- 2s2p $^3$P$^o_2$) forbidden transitions of Ti XIX.}
\label{fig:5}       
\end{figure*}
\clearpage


\begin{figure*}
\includegraphics[scale=0.70,angle=90]{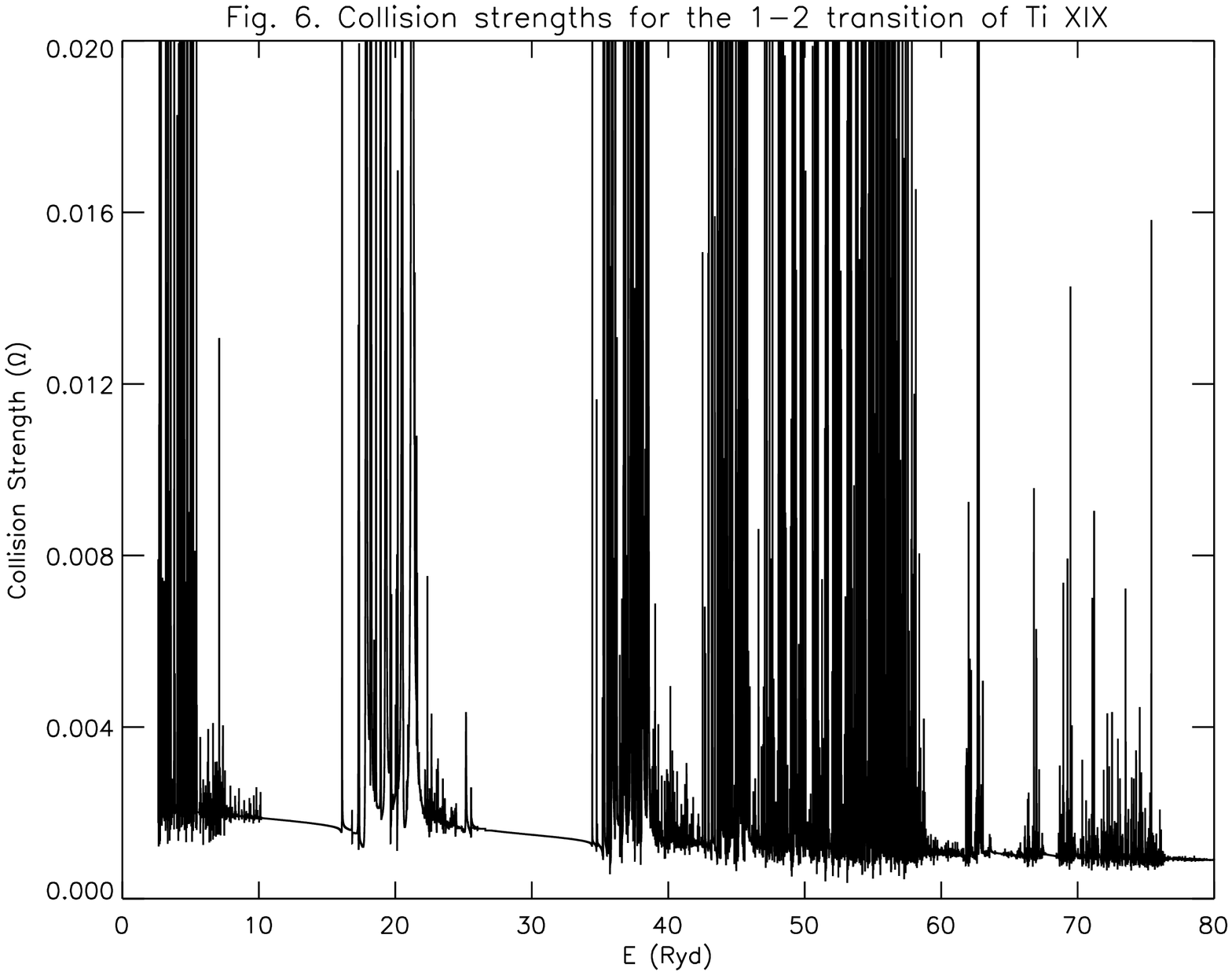}
\caption{Collision strengths for the 2s$^2$ $^1$S$_0$ - 2s2p $^3$P$^o_0$   (1--2) transition of Ti XIX.}
\label{fig:7}       
\end{figure*}

\begin{figure*}
\includegraphics[scale=0.70,angle=90]{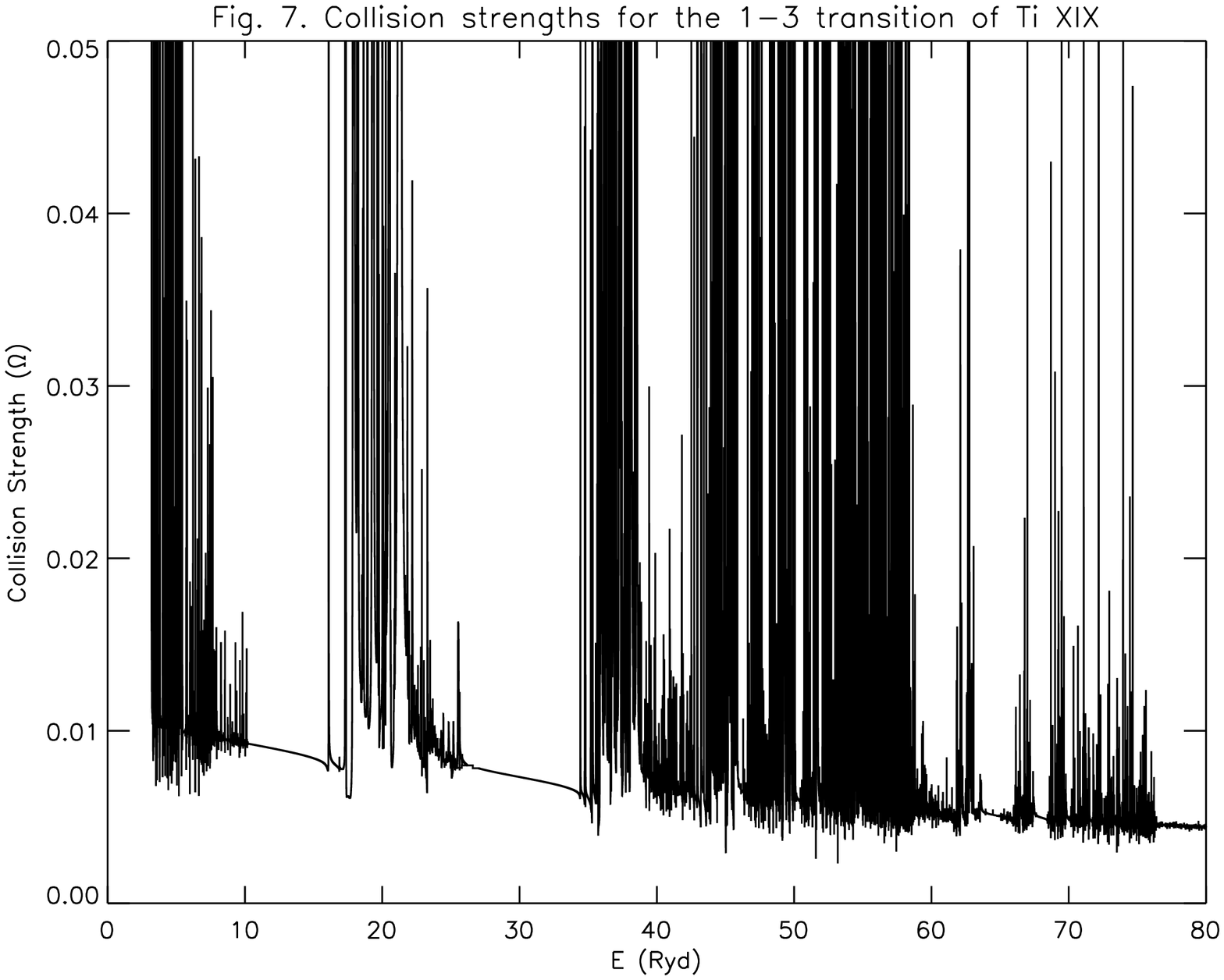}
\caption{Collision strengths for the  2s$^2$ $^1$S$_0$ - 2s2p $^3$P$^o_1$ (1--3) transition of Ti XIX.}
\label{fig:8}       
\end{figure*}

\begin{figure*}
\includegraphics[scale=0.70,angle=90]{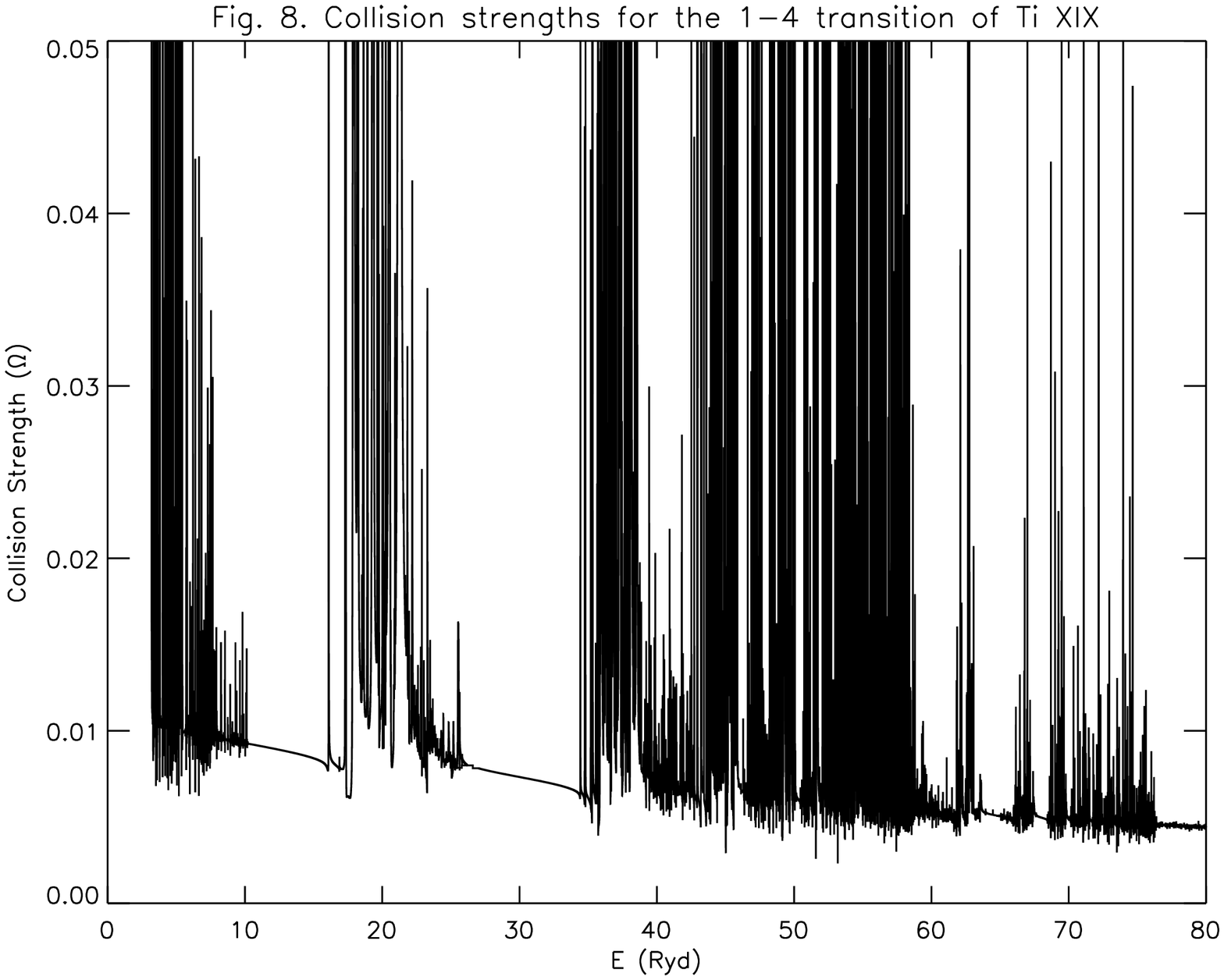}
\caption{Collision strengths for the 2s$^2$ $^1$S$_0$ - 2s2p $^3$P$^o_2$ (1--4) transition of Ti XIX.}
\label{fig:9}       
\end{figure*}

\begin{figure*}
\includegraphics[scale=0.70,angle=90]{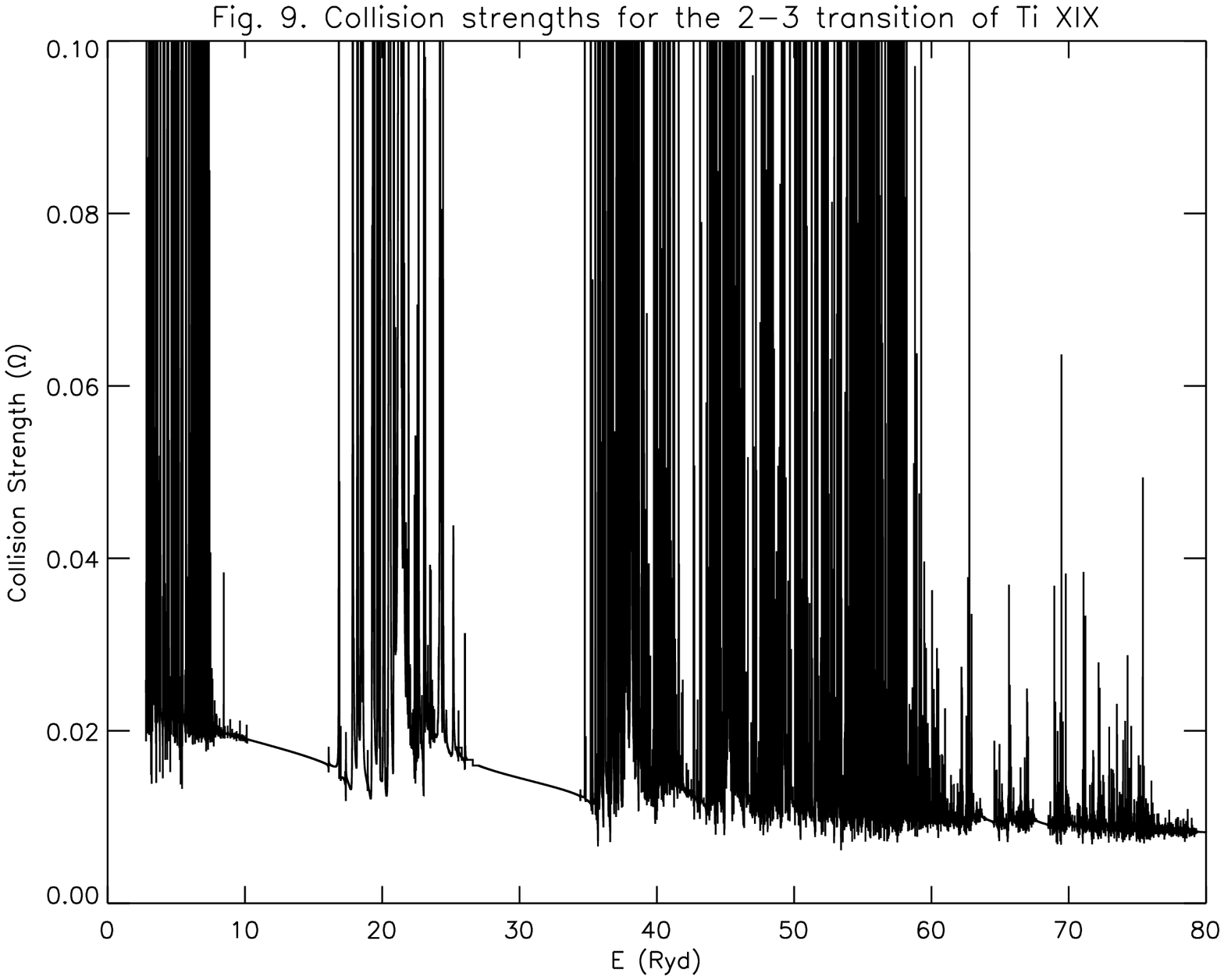}
\caption{Collision strengths for the  2s2p $^3$P$^o_0$  - 2s2p $^3$P$^o_1$ (2--3) transition of Ti XIX.}
\label{fig:10}       
\end{figure*}

\setcounter{figure} {9}
\begin{figure*}
\includegraphics[scale=0.70,angle=90]{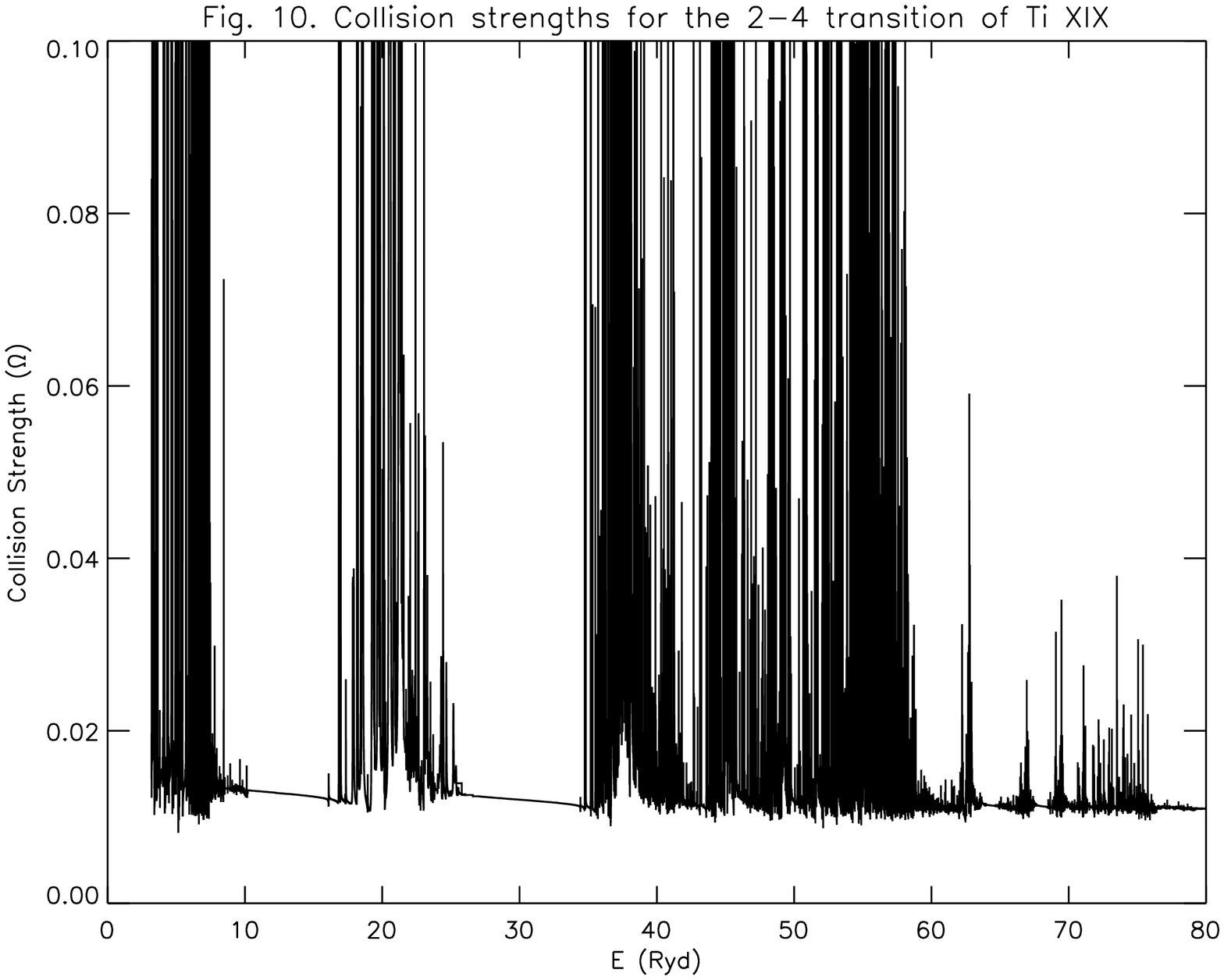}
\caption{Collision strengths for the  2s2p $^3$P$^o_0$  -  2s2p $^3$P$^o_2$  (2--4) transition of Ti XIX.}
\label{fig:10}       

\end{figure*}

\setcounter{figure} {10}
\begin{figure*}
\includegraphics[scale=0.70,angle=90]{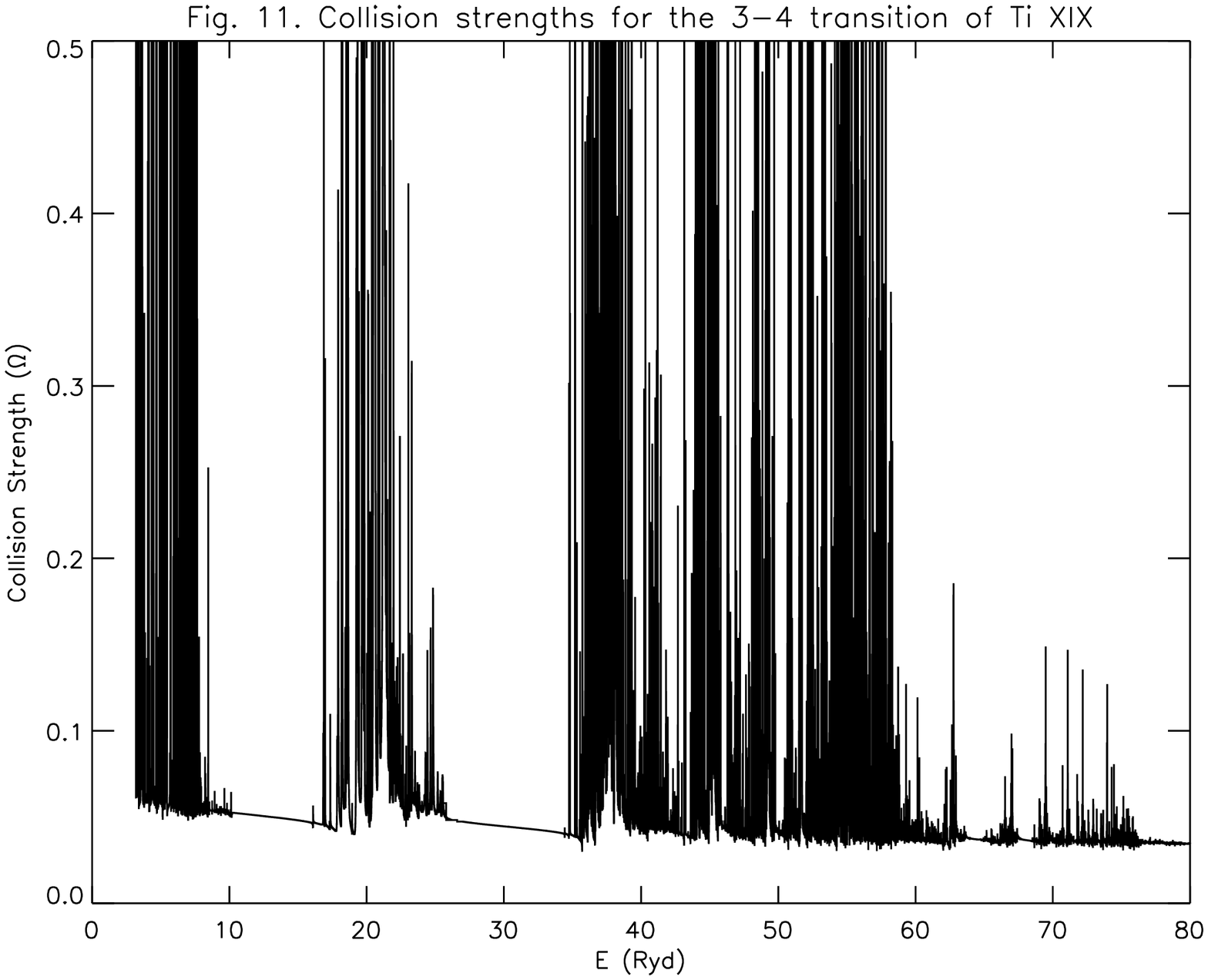}
\caption{Collision strengths for the  2s2p $^3$P$^o_1$  - 2s2p $^3$P$^o_2$  (3--4) transition of Ti XIX.}
\label{fig:10}       

\end{figure*}


\clearpage

\begin{table*} 
\caption{Energy levels (in Ryd) of Ti XIX and their lifetimes ($\tau$, s). $a{\pm}b \equiv a{\times}$10$^{{\pm}b}$.} 
\begin{tabular}{rllrrrrrl} \hline
Index  & Configuration       & Level     &  NIST   &   GRASP1 & GRASP2  &  FAC1   & FAC2   & $\tau$ (s$^{-1}$) \\ 
\hline
   1 &  2s$^2$    &  $^1$S$  _0$  &   0.00000	  &  	0.00000  &    0.00000	&  0.00000   &   0.00000   &   ........    \\ 
   2 &  2s2p	  &  $^3$P$^o_0$  &   2.62618	  &  	2.62268  &    2.63105	&  2.63967   &   2.63907   &   ........    \\ 
   3 &  2s2p	  &  $^3$P$^o_1$  &   2.77590	  &  	2.78404  &    2.78119	&  2.78920   &   2.78859   &   7.450-08    \\ 
   4 &  2s2p	  &  $^3$P$^o_2$  &   3.16447	  &  	3.18796  &    3.16788	&  3.17463   &   3.17411   &   9.798-04    \\ 
   5 &  2s2p	  &  $^1$P$^o_1$  &   5.37367	  &  	5.46387  &    5.45302	&  5.45146   &   5.44666   &   7.155-11    \\ 
   6 &  2p$^2$    &  $^3$P$  _0$  &   7.06970	  &  	7.10469  &    7.10937	&  7.12758   &   7.12669   &   1.040-10    \\ 
   7 &  2p$^2$    &  $^3$P$  _1$  &   7.33470	  &  	7.37764  &    7.37028	&  7.38761   &   7.38685   &   9.398-11    \\ 
   8 &  2p$^2$    &  $^3$P$  _2$  &   7.58548	  &  	7.65312  &    7.62549	&  7.64194   &   7.64085   &   9.595-11    \\ 
   9 &  2p$^2$    &  $^1$D$  _2$  &   8.36160	  &  	8.46555  &    8.43606	&  8.45099   &   8.44729   &   2.281-10    \\ 
  10 &  2p$^2$    &  $^1$S$  _0$  &  10.06194	  &    10.20533  &   10.19858	& 10.20719   &  10.20455   &   4.510-11    \\ 
  11 &  2s3s	  &  $^3$S$  _1$  &  56.14134	  &    56.22065  &   56.18639	& 56.18510   &  56.18496   &   5.249-13    \\ 
  12 &  2s3s	  &  $^1$S$  _0$  &  		  &    56.72694  &   56.69454	& 56.70758   &  56.70737   &   1.471-12    \\
  13 &  2s3p	  &  $^3$P$^o_1$  &  57.43898	  &    57.45953  &   57.42614	& 57.43735   &  57.43687   &   3.601-13    \\ 
  14 &  2s3p	  &  $^3$P$^o_0$  &  		  &    57.46676  &   57.43697	& 57.44853   &  57.44867   &   3.773-11    \\
  15 &  2s3p	  &  $^1$P$^o_1$  &  		  &    57.59804  &   57.55941	& 57.57007   &  57.56953   &   3.042-13    \\
  16 &  2s3p	  &  $^3$P$^o_2$  &  		  &    57.62728  &   57.58931	& 57.59959   &  57.59974   &   2.674-11    \\
  17 &  2s3d	  &  $^3$D$  _1$  &  58.14431	  &    58.29311  &   58.25119	& 58.26517   &  58.26290   &   9.053-14    \\  
  18 &  2s3d	  &  $^3$D$  _2$  &  58.22267	  &    58.31519  &   58.27088	& 58.28462   &  58.28231   &   9.184-14    \\ 
  19 &  2s3d	  &  $^3$D$  _3$  &  58.34570	  &    58.34899  &   58.30331	& 58.31672   &  58.31437   &   9.354-14    \\ 
  20 &  2s3d	  &  $^1$D$  _2$  &  58.73936	  &    58.84337  &   58.80058	& 58.81335   &  58.80922   &   1.310-13    \\ 
  21 &  2p3s	  &  $^3$P$^o_0$  &  		  &    59.52609  &   59.49725	& 59.52364   &  59.52383   &   7.179-13    \\
  22 &  2p3s	  &  $^3$P$^o_1$  &  		  &    59.63458  &   59.60204	& 59.62881   &  59.62857   &   6.572-13    \\
  23 &  2p3s	  &  $^3$P$^o_2$  &  		  &    60.10117  &   60.04739	& 60.07168   &  60.07186   &   6.185-13    \\
  24 &  2p3p	  &  $^3$D$  _1$  &  		  &    60.35157  &   60.32192	& 60.34961   &  60.34956   &   4.444-13    \\
  25 &  2p3s	  &  $^1$P$^o_1$  &  		  &    60.50059  &   60.45115	& 60.48131   &  60.47781   &   5.018-13    \\
  26 &  2p3p	  &  $^3$D$  _2$  &  60.66305	  &    60.69920  &   60.65945	& 60.68657   &  60.68621   &   5.143-13    \\ 
  27 &  2p3p	  &  $^1$P$  _1$  &  60.66305	  &    60.70962  &   60.67058	& 60.69893   &  60.69880   &   3.752-13    \\ 
  28 &  2p3p	  &  $^3$P$  _0$  &  60.92549	  &    60.97659  &   60.94744	& 60.99706   &  60.99525   &   3.291-13    \\ 
  29 &  2p3p	  &  $^3$D$  _3$  &  61.05216	  &    61.12956  &   61.06979	& 61.09317   &  61.09299   &   5.181-13    \\ 
  30 &  2p3p	  &  $^3$P$  _1$  &  		  &    61.15874  &   61.10785	& 61.13882   &  61.13841   &   3.461-13    \\
  31 &  2p3d	  &  $^3$F$^o_2$  &  		  &    61.17654  &   61.13912	& 61.17442   &  61.17252   &   8.358-13    \\
  32 &  2p3p	  &  $^3$S$  _1$  &  61.06674	  &    61.33556  &   61.28482	& 61.32152   &  61.32070   &   3.430-13    \\ 
  33 &  2p3p	  &  $^3$P$  _2$  &  		  &    61.39848  &   61.34579	& 61.39096   &  61.38959   &   3.161-13    \\
  34 &  2p3d	  &  $^3$F$^o_3$  &  		  &    61.40858  &   61.36396	& 61.40354   &  61.40123   &   5.417-13    \\
  35 &  2p3d	  &  $^1$D$^o_2$  &  61.40117	  &    61.51326  &   61.46906	& 61.50920   &  61.50879   &   1.588-13    \\ 
  36 &  2p3d	  &  $^3$D$^o_1$  &  61.58343	  &    61.70601  &   61.66333	& 61.70124   &  61.70128   &   7.765-14    \\ 
  37 &  2p3p	  &  $^1$D$  _2$  &  61.70098	  &    61.79050  &   61.73569	& 61.78649   &  61.78290   &   2.429-13    \\ 
  38 &  2p3d	  &  $^3$F$^o_4$  &  		  &    61.79095  &   61.72657	& 61.76010   &  61.75762   &   1.967-10    \\
  39 &  2p3d	  &  $^3$D$^o_2$  &  61.86683	  &    61.92348  &   61.86548	& 61.90434   &  61.90414   &   1.099-13    \\ 
  40 &  2p3d	  &  $^3$D$^o_3$  &  62.03542	  &    62.09265  &   62.02908	& 62.06790   &  62.06775   &   7.894-14    \\ 
  41 &  2p3d	  &  $^3$P$^o_2$  &  62.09100	  &    62.22337  &   62.16055	& 62.19689   &  62.19686   &   1.048-13    \\ 
  42 &  2p3d	  &  $^3$P$^o_1$  &  62.09100	  &    62.24732  &   62.18674	& 62.22404   &  62.22392   &   1.151-13    \\ 
  43 &  2p3d	  &  $^3$P$^o_0$  &  		  &    62.26794  &   62.21350	& 62.25042   &  62.25021   &   1.319-13    \\
  44 &  2p3p	  &  $^1$S$  _0$  &  		  &    62.40839  &   62.36301	& 62.41988   &  62.40815   &   3.455-13    \\
  45 &  2p3d	  &  $^1$F$^o_3$  &  62.61954	  &    62.71188  &   62.64738	& 62.68766   &  62.68128   &   5.713-14    \\ 
  46 &  2p3d	  &  $^1$P$^o_1$  &  62.56760	  &    62.77356  &   62.71547	& 62.75247   &  62.75037   &   9.437-14    \\   
\hline 													     
\end{tabular}      
\end{table*}

\setcounter{table}{0} 
\begin{table*} 
\caption{Energy levels (in Ryd) of Ti XIX and their lifetimes ($\tau$, s). $a{\pm}b \equiv a{\times}$10$^{{\pm}b}$.} 
\begin{tabular}{rllrrrrrl} \hline
Index  & Configuration       & Level     &  NIST   &   GRASP1 & GRASP2  &  FAC1   & FAC2   & $\tau$ (s$^{-1}$) \\ 
\hline
  47 &  2s4s	  &  $^3$S$  _1$  &  		  &    75.38981  &   75.34908	& 75.35360   &  75.35248   &   9.907-13    \\
  48 &  2s4s	  &  $^1$S$  _0$  &  		  &    75.56940  &   75.52980	& 75.53766   &  75.53503   &   1.225-12    \\
  49 &  2s4p	  &  $^3$P$^o_0$  &  	        &   75.87605	&   75.83721   & 75.84719   &  75.84729   &   2.295-12   \\
  50 &  2s4p	  &  $^3$P$^o_1$  &  	        &   75.88694	&   75.84697   & 75.85715   &  75.85692   &   1.260-12   \\
  51 &  2s4p	  &  $^3$P$^o_2$  &  	        &   75.94341	&   75.90115   & 75.91045   &  75.91059   &   2.368-12   \\
  52 &  2s4p	  &  $^1$P$^o_1$  &  75.87574   &   75.98058	&   75.93788   & 75.94922   &  75.94696   &   3.389-13   \\ 
  53 &  2s4d	  &  $^3$D$  _1$  &  76.24025   &   76.21957	&   76.17599   & 76.18372   &  76.18259   &   2.312-13   \\ 
  54 &  2s4d	  &  $^3$D$  _2$  &  76.20562   &   76.22736	&   76.18294   & 76.19065   &  76.18951   &   2.328-13   \\ 
  55 &  2s4d	  &  $^3$D$  _3$  &  76.18284   &   76.24023	&   76.19533   & 76.20300   &  76.20187   &   2.351-13   \\ 
  56 &  2s4d	  &  $^1$D$  _2$  &  	        &   76.39832	&   76.35435   & 76.36012   &  76.35793   &   2.616-13   \\
  57 &  2s4f	  &  $^3$F$^o_2$  &  	        &   76.41805	&   76.37418   & 76.38062   &  76.37811   &   5.349-13   \\
  58 &  2s4f	  &  $^3$F$^o_3$  &  	        &   76.42175	&   76.37711   & 76.38354   &  76.38100   &   5.353-13   \\
  59 &  2s4f	  &  $^3$F$^o_4$  &  	        &   76.42768	&   76.38273   & 76.38910   &  76.38654   &   5.359-13   \\
  60 &  2s4f	  &  $^1$F$^o_3$  &  	        &   76.46613	&   76.42163   & 76.42907   &  76.42635   &   5.456-13   \\
  61 &  2p4s	  &  $^3$P$^o_0$  &  	        &   78.45500	&   78.41892   & 78.44767   &  78.44781   &   1.199-12   \\
  62 &  2p4s	  &  $^3$P$^o_1$  &  	        &   78.49239	&   78.45542   & 78.48484   &  78.48335   &   1.034-12   \\
  63 &  2p4p	  &  $^3$D$  _1$  &  	        &   78.81310	&   78.77651   & 78.80826   &  78.80833   &   5.671-13   \\
  64 &  2p4p	  &  $^3$P$  _1$  &  	        &   78.97047	&   78.93198   & 78.96632   &  78.96487   &   5.426-13   \\
  65 &  2p4p	  &  $^3$D$  _2$  &  	        &   78.97802	&   78.93780   & 78.97186   &  78.97107   &   5.830-13   \\
  66 &  2p4s	  &  $^3$P$^o_2$  &  	        &   79.03315	&   78.97319   & 78.99945   &  78.99966   &   8.714-13   \\
  67 &  2p4p	  &  $^3$P$  _0$  &  	        &   79.04384	&   79.00783   & 79.04921   &  79.04442   &   5.681-13   \\
  68 &  2p4s	  &  $^1$P$^o_1$  &  	        &   79.12885	&   79.06882   & 79.09698   &  79.09180   &   8.234-13   \\
  69 &  2p4d	  &  $^3$F$^o_2$  &  	        &   79.15188	&   79.11203   & 79.14333   &  79.14216   &   7.316-13   \\
  70 &  2p4d	  &  $^3$F$^o_3$  &  	        &   79.27787	&   79.23583   & 79.26668   &  79.26493   &   3.510-13   \\
  71 &  2p4d	  &  $^3$D$^o_2$  &  	        &   79.28651	&   79.24524   & 79.27575   &  79.27499   &   2.728-13   \\
  72 &  2p4d	  &  $^3$D$^o_1$  &  	        &   79.34204	&   79.30120   & 79.33063   &  79.32966   &   1.920-13   \\
  73 &  2p4f	  &  $^3$G$  _3$  &  	        &   79.37004	&   79.32910   & 79.35644   &  79.35351   &   5.400-13   \\
  74 &  2p4f	  &  $^3$F$  _2$  &  	        &   79.39082	&   79.34979   & 79.37766   &  79.37785   &   5.433-13   \\
  75 &  2p4f	  &  $^3$F$  _3$  &  	        &   79.39114	&   79.35006   & 79.37781   &  79.37792   &   5.395-13   \\
  76 &  2p4f	  &  $^3$G$  _4$  &  	        &   79.39525	&   79.35418   & 79.38193   &  79.37885   &   5.535-13   \\
  77 &  2p4p	  &  $^1$P$  _1$  &  	        &   79.45185	&   79.39094   & 79.42149   &  79.42107   &   5.108-13   \\
  78 &  2p4p	  &  $^3$D$  _3$  &  79.36954   &   79.47881	&   79.41565   & 79.44415   &  79.44431   &   6.212-13   \\ 
  79 &  2p4p	  &  $^3$P$  _2$  &  	        &   79.52313	&   79.46271   & 79.49773   &  79.49632   &   5.502-13   \\
  80 &  2p4p	  &  $^3$S$  _1$  &  	        &   79.53841	&   79.47775   & 79.51011   &  79.50674   &   5.276-13   \\
  81 &  2p4p	  &  $^1$D$  _2$  &  	        &   79.67783	&   79.61622   & 79.65479   &  79.65120   &   4.909-13   \\
  82 &  2p4d	  &  $^3$F$^o_4$  &  	        &   79.75433	&   79.68940   & 79.71828   &  79.71673   &   1.088-12   \\
  83 &  2p4d	  &  $^1$D$^o_2$  &  79.70306   &   79.76625	&   79.70271   & 79.73117   &  79.73051   &   3.384-13   \\ 
  84 &  2p4d	  &  $^3$D$^o_3$  &  	        &   79.82750	&   79.76291   & 79.79074   &  79.79028   &   2.260-13   \\
  85 &  2p4d	  &  $^3$P$^o_2$  &  79.46887   &   79.88056	&   79.81625   & 79.84389   &  79.84298   &   2.290-13   \\ 
  86 &  2p4d	  &  $^3$P$^o_1$  &  	        &   79.88919	&   79.82568   & 79.85354   &  79.85212   &   2.431-13   \\
  87 &  2p4d	  &  $^3$P$^o_0$  &  	        &   79.89819	&   79.83705   & 79.86505   &  79.86304   &   2.797-13   \\
  88 &  2p4p	  &  $^1$S$  _0$  &  	        &   79.90379	&   79.84573   & 79.89239   &  79.88819   &   6.549-13   \\
 \hline
\end{tabular}	  
\end{table*} 
  
\setcounter{table}{0} 
\begin{table*} 
\caption{Energy levels (in Ryd) of Ti XIX and their lifetimes ($\tau$, s). $a{\pm}b \equiv a{\times}$10$^{{\pm}b}$.} 
\begin{tabular}{rllrrrrrl} \hline
Index  & Configuration       & Level     &  NIST   &   GRASP1 & GRASP2  &  FAC1   & FAC2   & $\tau$ (s$^{-1}$) \\ 
\hline  
  89 &  2p4f	  &  $^1$F$  _3$  &  	        &   79.92931	&   79.86466   & 79.88954   &  79.86857   &   5.386-13   \\
  90 &  2p4f	  &  $^3$F$  _4$  &  	        &   79.94410	&   79.87919   & 79.90411   &  79.90253   &   5.439-13   \\
  91 &  2p4f	  &  $^3$D$  _3$  &  	        &   79.97863	&   79.91434   & 79.93935   &  79.93950   &   5.380-13   \\
  92 &  2p4f	  &  $^3$D$  _2$  &  	        &   79.98128	&   79.91719   & 79.94263   &  79.94099   &   5.385-13   \\
  93 &  2p4f	  &  $^3$G$  _5$  &  	        &   79.98694	&   79.92135   & 79.94597   &  79.94267   &   5.429-13   \\
  94 &  2p4f	  &  $^1$G$  _4$  &  	        &   80.00956	&   79.94390   & 79.96957   &  79.96434   &   5.681-13   \\
  95 &  2p4f	  &  $^3$D$  _1$  &  	        &   80.03020	&   79.96676   & 79.99200   &  79.99212   &   5.353-13   \\
  96 &  2p4d	  &  $^1$F$^o_3$  &  79.96824   &   80.04175	&   79.97704   & 80.00288   &  79.99691   &   1.416-13   \\ 
  97 &  2p4f	  &  $^1$D$  _2$  &  	        &   80.05501	&   79.99117   & 80.01690   &  80.01675   &   5.388-13   \\
  98 &  2p4d	  &  $^1$P$^o_1$  &  	        &   80.07038	&   80.00818   & 80.03399   &  80.02969   &   2.077-13   \\
 \hline	
\end{tabular}			      
\begin{flushleft}
{\small
NIST: {\tt http://nist.gov/pml/data/asd.cfm} \\
GRASP1: Energies from the {\sc grasp} code with 98 level calculations {\em without} Breit and QED effects \\
GRASP2: Energies from the {\sc grasp} code with 98 level calculations {\em with} Breit and QED effects \\
FAC1: Energies from the {\sc fac} code with 98 level calculations \\
FAC2: Energies from the {\sc fac} code with 166 level calculations \\
}
\end{flushleft}
\end{table*} 

\clearpage

\setcounter{table}{2} 

\begin{table*}                                                                                                
\caption{Comparison between GRASP and FAC  A- values (s$^{-1}$) for some transitions of  Ti XIX.  ($a{\pm}b \equiv$ $a\times$10$^{{\pm}b}$).}       
\begin{tabular}{rrlllllll}                                                                                                                                                                                          
\hline    
  $i$ & $j$ &   f (GRASP)    &    A (GRASP) & A (FAC)    & A(GRASP)/A(FAC)      \\					       
\hline                                                                                                      
     1 &    3 &   6.4812$-$04 &      1.3423$+$07 &   1.359$+$07 &     0.99 \\ 
     1 &    5 &   1.7554$-$01 &      1.3976$+$10 &   1.395$+$10 &     1.00 \\ 
     1 &   13 &   3.0051$-$01 &      2.6535$+$12 &   2.704$+$12 &     0.98 \\ 
     1 &   15 &   3.5360$-$01 &      3.1367$+$12 &   3.139$+$12 &     1.00 \\ 
     2 &    7 &   6.9804$-$02 &      4.1978$+$09 &   4.216$+$09 &     1.00 \\ 
     2 &   11 &   2.7373$-$02 &      2.1021$+$11 &   2.138$+$11 &     0.98 \\ 
     2 &   17 &   7.4379$-$01 &      6.1609$+$12 &   6.164$+$12 &     1.00 \\ 
      3 &    7 &   1.6691$-$02 &      2.8235$+$09 &   2.836$+$09 &     1.00 \\ 
     3 &    8 &   2.9978$-$02 &      3.3905$+$09 &   3.406$+$09 &     1.00 \\ 
     3 &    9 &   7.9155$-$04 &      1.2199$+$08 &   1.216$+$08 &     1.00 \\ 
     3 &   10 &   5.0895$-$05 &      6.7477$+$07 &   6.728$+$07 &     1.00 \\ 
     3 &   11 &   2.7504$-$02 &      6.3011$+$11 &   6.398$+$11 &     0.98 \\ 
     3 &   12 &   3.2644$-$05 &      2.2865$+$09 &   2.542$+$09 &     0.90 \\ 
     3 &   17 &   1.8421$-$01 &      4.5527$+$12 &   4.552$+$12 &     1.00 \\ 
     3 &   18 &   5.5139$-$01 &      8.1824$+$12 &   8.191$+$12 &     1.00 \\ 
     3 &   20 &   1.3291$-$03 &      2.0101$+$10 &   2.005$+$10 &     1.00 \\ 
     4 &    7 &   1.5298$-$02 &      3.6169$+$09 &   3.639$+$09 &     0.99 \\ 
     4 &    8 &   4.3448$-$02 &      6.9346$+$09 &   6.970$+$09 &     0.99 \\ 
     4 &    9 &   6.6129$-$03 &      1.4742$+$09 &   1.481$+$09 &     1.00 \\ 
     4 &   11 &   2.8159$-$02 &      1.0597$+$12 &   1.073$+$12 &     0.99 \\ 
     4 &   17 &   7.3864$-$03 &      3.0003$+$11 &   2.995$+$11 &     1.00 \\ 
     4 &   18 &   1.1007$-$01 &      2.6845$+$12 &   2.684$+$12 &     1.00 \\ 
     4 &   19 &   6.1290$-$01 &      1.0690$+$13 &   1.071$+$13 &     1.00 \\ 
     4 &   20 &   1.2753$-$04 &      3.1706$+$09 &   3.292$+$09 &     0.96 \\ 
     5 &    6 &   1.5967$-$04 &      1.0556$+$07 &   1.086$+$07 &     0.97 \\ 
     5 &    7 &   6.2606$-$05 &      1.8485$+$06 &   1.906$+$06 &     0.97 \\ 
     5 &    8 &   4.2613$-$03 &      9.6929$+$07 &   9.917$+$07 &     0.98 \\ 
     5 &    9 &   6.5004$-$02 &      2.7878$+$09 &   2.832$+$09 &     0.98 \\ 
     5 &   10 &   4.0732$-$02 &      2.2105$+$10 &   2.220$+$10 &     1.00 \\ 
     5 &   11 &   2.4577$-$04 &      5.0813$+$09 &   5.156$+$09 &     0.99 \\ 
     5 &   12 &   1.0706$-$02 &      6.7738$+$11 &   6.918$+$11 &     0.98 \\ 
     5 &   17 &   1.4504$-$03 &      3.2478$+$10 &   3.240$+$10 &     1.00 \\ 
     5 &   18 &   1.6119$-$03 &      2.1672$+$10 &   2.255$+$10 &     0.96 \\ 
     5 &   20 &   5.5415$-$01 &      7.6008$+$12 &   7.577$+$12 &     1.00 \\ 
     6 &   13 &   6.4543$-$04 &      4.3753$+$09 &   4.612$+$09 &     0.95 \\ 
     6 &   15 &   8.4843$-$04 &      5.7819$+$09 &   5.845$+$09 &     0.99 \\ 
     7 &   13 &   1.9886$-$04 &      4.0023$+$09 &   4.093$+$09 &     0.98 \\ 
     7 &   14 &   3.9853$-$04 &      2.4073$+$10 &   2.495$+$10 &     0.96 \\ 
     7 &   15 &   7.6560$-$05 &      1.5491$+$09 &   1.648$+$09 &     0.94 \\ 
     7 &   16 &   9.2907$-$04 &      1.1292$+$10 &   1.149$+$10 &     0.98 \\ 
     8 &   13 &   1.1242$-$03 &      3.7325$+$10 &   3.847$+$10 &     0.97 \\ 
     8 &   15 &   1.7751$-$04 &      5.9255$+$09 &   5.804$+$09 &     1.02 \\ 
     8 &   16 &   1.0792$-$03 &      2.1640$+$10 &   2.213$+$10 &     0.98 \\ 
     9 &   13 &   2.1883$-$03 &      7.0310$+$10 &   7.267$+$10 &     0.97 \\ 
     9 &   15 &   4.0870$-$03 &      1.3203$+$11 &   1.333$+$11 &     0.99 \\ 
     9 &   16 &   8.8457$-$05 &      1.7167$+$09 &   1.798$+$09 &     0.95 \\ 
    10 &   13 &   9.6368$-$04 &      5.7551$+$09 &   5.311$+$09 &     1.08 \\ 
    10 &   15 &   5.1006$-$04 &      3.0633$+$09 &   2.620$+$09 &     1.17 \\ 
    11 &   13 &   2.8545$-$02 &      3.5242$+$08 &   3.576$+$08 &     0.99 \\ 
    11 &   14 &   1.7081$-$02 &      6.4376$+$08 &   6.595$+$08 &     0.98 \\ 
    11 &   15 &   2.4734$-$02 &      3.7454$+$08 &   3.864$+$08 &     0.97 \\ 
    11 &   16 &   9.6721$-$02 &      9.1747$+$08 &   9.343$+$08 &     0.98 \\ 

\hline                                                                                                        
\end{tabular}                                                                                                 
\end{table*}                                                                                                                                                    

\clearpage

\setcounter{table}{3} 
\begin{table*}                                                                                                
\caption{Collision strengths for resonance  transitions of  Ti XIX. ($a{\pm}b \equiv$ $a\times$10$^{{\pm}b}$).}       
\begin{tabular}{rrllllllll}                                                                                   
\hline                                                                                                        
\hline                                                                                                        
\multicolumn{2}{c}{Transition} & \multicolumn{7}{c}{Energy (Ryd)}\\                                           
\hline                                                                                                        
  $i$ & $j$ &    100 &   300 &   500 &  700  & 900  &  1100 &  FAC$^c$ \\                                         
\hline                                                                                                        
  1 &  2 &  7.561$-$4 &  2.120$-$4 &  9.758$-$5 &  5.602$-$5 &  3.634$-$5 &  2.552$-$5 &  6.199$-$5 \\
  1 &  3 &  8.260$-$3 &  6.955$-$3 &  6.406$-$3 &  5.839$-$3 &  5.803$-$3 &  5.396$-$3 &  8.643$-$3 \\
  1 &  4 &  3.693$-$3 &  1.029$-$3 &  4.723$-$4 &  2.708$-$4 &  1.755$-$4 &  1.232$-$4 &  3.004$-$4 \\
  1 &  5 &  6.500$-$1 &  7.992$-$1 &  8.038$-$1 &  7.428$-$1 &  7.618$-$1 &  6.805$-$1 &  9.797$-$1 \\
  1 &  6 &  7.543$-$5 &  5.139$-$5 &  4.662$-$5 &  4.490$-$5 &  4.415$-$5 &  4.381$-$5 &  3.522$-$5 \\
  1 &  7 &  6.142$-$5 &  1.110$-$5 &  3.823$-$6 &  1.763$-$6 &  9.633$-$7 &  5.860$-$7 &  2.136$-$6 \\
  1 &  8 &  3.899$-$4 &  3.753$-$4 &  3.855$-$4 &  3.946$-$4 &  4.008$-$4 &  4.053$-$4 &  3.794$-$4 \\
  1 &  9 &  1.956$-$3 &  2.251$-$3 &  2.366$-$3 &  2.436$-$3 &  2.480$-$3 &  2.510$-$3 &  2.516$-$3 \\
  1 & 10 &  9.304$-$4 &  7.814$-$4 &  7.271$-$4 &  7.010$-$4 &  6.867$-$4 &  6.788$-$4 &  6.010$-$4 \\
  1 & 11 &  7.696$-$4 &  1.533$-$4 &  6.489$-$5 &  3.589$-$5 &  2.281$-$5 &  1.587$-$5 &  3.429$-$5 \\
  1 & 12 &  2.021$-$2 &  2.373$-$2 &  2.471$-$2 &  2.526$-$2 &  2.568$-$2 &  2.605$-$2 &  2.502$-$2 \\
  1 & 13 &  9.225$-$3 &  2.333$-$2 &  3.210$-$2 &  3.855$-$2 &  4.390$-$2 &  4.737$-$2 &  4.099$-$2 \\
  1 & 14 &  2.251$-$4 &  3.678$-$5 &  1.397$-$5 &  7.212$-$6 &  4.372$-$6 &  2.925$-$6 &  6.113$-$6 \\
  1 & 15 &  1.093$-$2 &  2.841$-$2 &  3.923$-$2 &  4.718$-$2 &  5.381$-$2 &  5.802$-$2 &  4.675$-$2 \\
  1 & 16 &  1.107$-$3 &  1.807$-$4 &  6.856$-$5 &  3.537$-$5 &  2.143$-$5 &  1.433$-$5 &  3.005$-$5 \\
  1 & 17 &  1.212$-$3 &  2.005$-$4 &  7.732$-$5 &  4.038$-$5 &  2.466$-$5 &  1.659$-$5 &  3.809$-$5 \\
  1 & 18 &  2.070$-$3 &  4.163$-$4 &  2.215$-$4 &  1.651$-$4 &  1.422$-$4 &  1.304$-$4 &  1.726$-$4 \\
  1 & 19 &  2.835$-$3 &  4.686$-$4 &  1.807$-$4 &  9.434$-$5 &  5.761$-$5 &  3.874$-$5 &  8.936$-$5 \\
  1 & 20 &  3.295$-$2 &  5.466$-$2 &  6.168$-$2 &  6.518$-$2 &  6.732$-$2 &  6.843$-$2 &  6.551$-$2 \\
  1 & 21 &  3.380$-$6 &  6.468$-$7 &  2.543$-$7 &  1.332$-$7 &  8.108$-$8 &  5.445$-$8 &  1.309$-$7 \\
  1 & 22 &  2.104$-$4 &  5.242$-$4 &  7.223$-$4 &  8.692$-$4 &  9.903$-$4 &  1.068$-$3 &  8.388$-$4 \\
  1 & 23 &  1.464$-$5 &  2.855$-$6 &  1.129$-$6 &  5.942$-$7 &  3.637$-$7 &  2.453$-$7 &  5.739$-$7 \\
  1 & 24 &  3.916$-$5 &  9.693$-$6 &  4.779$-$6 &  3.012$-$6 &  2.147$-$6 &  1.648$-$6 &  1.402$-$6 \\
  1 & 25 &  7.582$-$4 &  1.930$-$3 &  2.663$-$3 &  3.208$-$3 &  3.655$-$3 &  3.945$-$3 &  3.425$-$3 \\
  1 & 26 &  8.262$-$5 &  5.163$-$5 &  4.967$-$5 &  4.994$-$5 &  5.051$-$5 &  5.094$-$5 &  4.393$-$5 \\
  1 & 27 &  2.660$-$5 &  7.211$-$6 &  3.857$-$6 &  2.572$-$6 &  1.907$-$6 &  1.506$-$6 &  8.167$-$7 \\
  1 & 28 &  1.064$-$5 &  6.978$-$6 &  6.648$-$6 &  6.608$-$6 &  6.627$-$6 &  6.671$-$6 &  6.480$-$6 \\
  1 & 29 &  6.875$-$5 &  1.412$-$5 &  5.666$-$6 &  2.985$-$6 &  1.824$-$6 &  1.226$-$6 &  3.074$-$6 \\
  1 & 30 &  1.825$-$5 &  4.900$-$6 &  2.644$-$6 &  1.778$-$6 &  1.326$-$6 &  1.052$-$6 &  5.392$-$7 \\
  1 & 31 &  4.376$-$5 &  7.752$-$6 &  4.091$-$6 &  2.927$-$6 &  2.353$-$6 &  2.004$-$6 &  8.870$-$7 \\
  1 & 32 &  1.424$-$5 &  2.587$-$6 &  1.006$-$6 &  5.299$-$7 &  3.279$-$7 &  2.233$-$7 &  4.067$-$7 \\
  1 & 33 &  6.018$-$5 &  7.790$-$5 &  8.599$-$5 &  9.012$-$5 &  9.268$-$5 &  9.422$-$5 &  8.231$-$5 \\
  1 & 34 &  6.250$-$5 &  2.166$-$5 &  1.869$-$5 &  1.832$-$5 &  1.843$-$5 &  1.863$-$5 &  1.714$-$5 \\
  1 & 35 &  4.534$-$5 &  1.091$-$5 &  7.120$-$6 &  5.685$-$6 &  4.861$-$6 &  4.298$-$6 &  8.240$-$7 \\
  1 & 36 &  2.527$-$4 &  4.217$-$4 &  5.235$-$4 &  5.984$-$4 &  6.596$-$4 &  6.996$-$4 &  5.493$-$4 \\
  1 & 37 &  1.927$-$4 &  3.241$-$4 &  3.639$-$4 &  3.824$-$4 &  3.933$-$4 &  3.999$-$4 &  3.693$-$4 \\
  1 & 38 &  6.185$-$5 &  8.327$-$6 &  3.009$-$6 &  1.534$-$6 &  9.289$-$7 &  6.233$-$7 &  1.475$-$6 \\
  1 & 39 &  3.274$-$5 &  7.004$-$6 &  4.295$-$6 &  3.339$-$6 &  2.817$-$6 &  2.472$-$6 &  6.715$-$7 \\
  1 & 40 &  2.451$-$5 &  8.938$-$6 &  8.289$-$6 &  8.386$-$6 &  8.565$-$6 &  8.732$-$6 &  8.455$-$6 \\
  1 & 41 &  5.272$-$5 &  8.133$-$6 &  3.189$-$6 &  1.742$-$6 &  1.127$-$6 &  8.066$-$7 &  1.496$-$6 \\
  1 & 42 &  5.086$-$5 &  2.868$-$5 &  3.061$-$5 &  3.360$-$5 &  3.646$-$5 &  3.838$-$5 &  3.233$-$5 \\
  1 & 43 &  1.476$-$5 &  2.298$-$6 &  8.707$-$7 &  4.513$-$7 &  2.749$-$7 &  1.847$-$7 &  4.387$-$7 \\
  1 & 44 &  7.843$-$5 &  7.521$-$5 &  7.541$-$5 &  7.593$-$5 &  7.651$-$5 &  7.717$-$5 &  7.792$-$5 \\
  1 & 45 &  1.622$-$4 &  2.208$-$4 &  2.399$-$4 &  2.514$-$4 &  2.598$-$4 &  2.661$-$4 &  2.609$-$4 \\
  1 & 46 &  1.534$-$3 &  2.721$-$3 &  3.392$-$3 &  3.883$-$3 &  4.283$-$3 &  4.545$-$3 &  3.815$-$3 \\
  1 & 47 &  3.390$-$4 &  5.704$-$5 &  2.317$-$5 &  1.257$-$5 &  7.906$-$6 &  5.458$-$6 &  1.158$-$5 \\
  1 & 48 &  3.738$-$3 &  4.659$-$3 &  4.909$-$3 &  5.042$-$3 &  5.141$-$3 &  5.224$-$3 &  5.034$-$3 \\
  1 & 49 &  1.157$-$4 &  1.581$-$5 &  5.686$-$6 &  2.854$-$6 &  1.702$-$6 &  1.124$-$6 &  2.476$-$6 \\
\hline                                                                                                        
\end{tabular}                                                                                                 
\end{table*}                                                                            

\clearpage  
\newpage 
\setcounter{table}{3} 
\begin{table*}                                                                                                
\caption{Collision strengths for resonance  transitions of  Ti XIX. ($a{\pm}b \equiv$ $a\times$10$^{{\pm}b}$).}     
\begin{tabular}{rrllllllll}                                                                                   
\hline                                                                                                        
\hline                                                                                                        
\multicolumn{2}{c}{Transition} & \multicolumn{7}{c}{Energy (Ryd)}\\                                           
\hline                                                                                                        
  $i$ & $j$ &    100 &   300 &   500 &  700  & 900  &  1100 &  FAC$^c$ \\                                         
\hline    
  1 & 50 &  7.954$-$4 &  1.383$-$3 &  1.864$-$3 &  2.225$-$3 &  2.518$-$3 &  2.725$-$3 &  2.222$-$3 \\                                                                                                    
  1 & 51 &  5.696$-$4 &  7.775$-$5 &  2.792$-$5 &  1.400$-$5 &  8.344$-$6 &  5.504$-$6 &  1.215$-$5 \\
  1 & 52 &  3.309$-$3 &  9.083$-$3 &  1.257$-$2 &  1.510$-$2 &  1.714$-$2 &  1.857$-$2 &  1.605$-$2 \\
  1 & 53 &  4.978$-$4 &  7.552$-$5 &  2.857$-$5 &  1.479$-$5 &  8.990$-$6 &  6.024$-$6 &  1.368$-$5 \\
  1 & 54 &  8.401$-$4 &  1.491$-$4 &  7.445$-$5 &  5.322$-$5 &  4.463$-$5 &  4.044$-$5 &  5.443$-$5 \\
  1 & 55 &  1.159$-$3 &  1.755$-$4 &  6.635$-$5 &  3.434$-$5 &  2.087$-$5 &  1.398$-$5 &  3.184$-$5 \\
  1 & 56 &  4.843$-$3 &  8.628$-$3 &  9.829$-$3 &  1.043$-$2 &  1.080$-$2 &  1.107$-$2 &  1.041$-$2 \\
  1 & 57 &  3.353$-$4 &  3.186$-$5 &  1.033$-$5 &  4.982$-$6 &  2.916$-$6 &  1.912$-$6 &  3.461$-$6 \\
  1 & 58 &  4.825$-$4 &  7.648$-$5 &  4.932$-$5 &  4.302$-$5 &  4.087$-$5 &  3.994$-$5 &  4.698$-$5 \\
  1 & 59 &  6.019$-$4 &  5.701$-$5 &  1.847$-$5 &  8.903$-$6 &  5.210$-$6 &  3.415$-$6 &  6.229$-$6 \\
  1 & 60 &  1.852$-$3 &  3.257$-$3 &  3.517$-$3 &  3.625$-$3 &  3.695$-$3 &  3.740$-$3 &  3.690$-$3 \\
  1 & 61 &  1.741$-$6 &  2.816$-$7 &  1.075$-$7 &  5.568$-$8 &  3.380$-$8 &  2.270$-$8 &  5.087$-$8 \\
  1 & 62 &  1.517$-$5 &  2.126$-$5 &  2.699$-$5 &  3.128$-$5 &  3.471$-$5 &  3.716$-$5 &  3.363$-$5 \\
  1 & 63 &  1.141$-$5 &  2.149$-$6 &  9.625$-$7 &  5.794$-$7 &  4.035$-$7 &  3.057$-$7 &  3.043$-$7 \\
  1 & 64 &  1.037$-$5 &  1.796$-$6 &  7.751$-$7 &  4.567$-$7 &  3.133$-$7 &  2.346$-$7 &  2.541$-$7 \\
  1 & 65 &  1.695$-$5 &  6.359$-$6 &  5.579$-$6 &  5.471$-$6 &  5.474$-$6 &  5.504$-$6 &  4.351$-$6 \\
  1 & 66 &  8.556$-$6 &  1.488$-$6 &  5.866$-$7 &  3.121$-$7 &  1.941$-$7 &  1.334$-$7 &  2.671$-$7 \\
  1 & 67 &  1.226$-$5 &  1.172$-$5 &  1.217$-$5 &  1.253$-$5 &  1.281$-$5 &  1.304$-$5 &  1.281$-$5 \\
  1 & 68 &  2.337$-$5 &  3.711$-$5 &  4.731$-$5 &  5.476$-$5 &  6.065$-$5 &  6.490$-$5 &  6.279$-$5 \\
  1 & 69 &  2.101$-$5 &  2.810$-$6 &  1.192$-$6 &  7.345$-$7 &  5.347$-$7 &  4.251$-$7 &  3.531$-$7 \\
  1 & 70 &  2.613$-$5 &  8.765$-$6 &  8.074$-$6 &  8.204$-$6 &  8.426$-$6 &  8.649$-$6 &  9.531$-$6 \\
  1 & 71 &  2.543$-$5 &  3.428$-$6 &  1.414$-$6 &  8.473$-$7 &  6.038$-$7 &  4.728$-$7 &  4.439$-$7 \\
  1 & 72 &  1.141$-$4 &  1.961$-$4 &  2.440$-$4 &  2.789$-$4 &  3.070$-$4 &  3.283$-$4 &  2.627$-$4 \\
  1 & 73 &  5.650$-$6 &  6.905$-$7 &  3.457$-$7 &  2.324$-$7 &  1.760$-$7 &  1.422$-$7 &  5.481$-$8 \\
  1 & 74 &  1.637$-$5 &  3.022$-$5 &  3.640$-$5 &  3.971$-$5 &  4.181$-$5 &  4.303$-$5 &  3.912$-$5 \\
  1 & 75 &  8.302$-$6 &  8.702$-$7 &  3.611$-$7 &  2.189$-$7 &  1.561$-$7 &  1.213$-$7 &  9.118$-$8 \\
  1 & 76 &  7.436$-$6 &  5.314$-$6 &  5.892$-$6 &  6.290$-$6 &  6.571$-$6 &  6.783$-$6 &  6.587$-$6 \\
  1 & 77 &  7.309$-$6 &  1.609$-$6 &  8.850$-$7 &  6.179$-$7 &  4.774$-$7 &  3.899$-$7 &  1.226$-$7 \\
  1 & 78 &  1.970$-$5 &  3.296$-$6 &  1.250$-$6 &  6.409$-$7 &  3.857$-$7 &  2.564$-$7 &  6.115$-$7 \\
  1 & 79 &  1.043$-$5 &  4.104$-$6 &  3.713$-$6 &  3.674$-$6 &  3.686$-$6 &  3.709$-$6 &  2.854$-$6 \\
  1 & 80 &  9.624$-$6 &  1.660$-$6 &  6.735$-$7 &  3.697$-$7 &  2.370$-$7 &  1.666$-$7 &  2.799$-$7 \\
  1 & 81 &  1.139$-$5 &  1.229$-$5 &  1.346$-$5 &  1.398$-$5 &  1.425$-$5 &  1.441$-$5 &  1.076$-$5 \\
  1 & 82 &  2.942$-$5 &  3.510$-$6 &  1.230$-$6 &  6.184$-$7 &  3.715$-$7 &  2.479$-$7 &  5.861$-$7 \\
  1 & 83 &  1.420$-$5 &  2.497$-$6 &  1.425$-$6 &  1.075$-$6 &  8.928$-$7 &  7.763$-$7 &  1.597$-$7 \\
  1 & 84 &  1.372$-$5 &  2.257$-$6 &  1.527$-$6 &  1.404$-$6 &  1.387$-$6 &  1.398$-$6 &  1.726$-$6 \\
  1 & 85 &  2.228$-$5 &  2.877$-$6 &  1.054$-$6 &  5.535$-$7 &  3.482$-$7 &  2.442$-$7 &  4.555$-$7 \\
  1 & 86 &  1.965$-$5 &  7.175$-$6 &  6.954$-$6 &  7.432$-$6 &  7.973$-$6 &  8.430$-$6 &  7.865$-$6 \\
  1 & 87 &  7.132$-$6 &  9.664$-$7 &  3.505$-$7 &  1.777$-$7 &  1.067$-$7 &  7.100$-$8 &  1.625$-$7 \\
  1 & 88 &  4.233$-$5 &  4.621$-$5 &  4.867$-$5 &  5.026$-$5 &  5.148$-$5 &  5.247$-$5 &  5.448$-$5 \\
  1 & 89 &  3.541$-$6 &  6.445$-$7 &  4.000$-$7 &  2.992$-$7 &  2.411$-$7 &  2.029$-$7 &  2.712$-$8 \\
  1 & 90 &  3.440$-$6 &  6.498$-$7 &  5.698$-$7 &  5.732$-$7 &  5.854$-$7 &  5.978$-$7 &  5.785$-$7 \\
  1 & 91 &  6.938$-$6 &  7.258$-$7 &  3.153$-$7 &  1.995$-$7 &  1.469$-$7 &  1.171$-$7 &  5.622$-$8 \\
  1 & 92 &  1.004$-$5 &  1.510$-$5 &  1.812$-$5 &  1.981$-$5 &  2.091$-$5 &  2.153$-$5 &  2.033$-$5 \\
  1 & 93 &  7.123$-$6 &  6.086$-$7 &  2.188$-$7 &  1.130$-$7 &  6.895$-$8 &  4.640$-$8 &  8.223$-$8 \\
  1 & 94 &  8.379$-$6 &  9.444$-$6 &  1.072$-$5 &  1.150$-$5 &  1.203$-$5 &  1.243$-$5 &  1.245$-$5 \\
  1 & 95 &  4.414$-$6 &  3.638$-$7 &  1.129$-$7 &  5.313$-$8 &  3.059$-$8 &  1.982$-$8 &  3.719$-$8 \\
  1 & 96 &  2.170$-$5 &  2.570$-$5 &  2.883$-$5 &  3.082$-$5 &  3.226$-$5 &  3.341$-$5 &  4.402$-$5 \\
  1 & 97 &  2.212$-$5 &  4.245$-$5 &  5.137$-$5 &  5.624$-$5 &  5.936$-$5 &  6.116$-$5 &  5.829$-$5 \\
  1 & 98 &  3.068$-$4 &  5.630$-$4 &  7.035$-$4 &  8.051$-$4 &  8.867$-$4 &  9.489$-$4 &  7.958$-$4 \\
\hline                                                                                                        
\end{tabular}                                                           
\begin{flushleft} 
$^c$: E $\sim$ 700 Ryd
\end{flushleft}                                                                                                 
\end{table*} 
                                                                
\clearpage      
\newpage                
\setcounter{table}{5}                                                                                         
\begin{table*}                                                                                                
\caption{Comparison of $\Upsilon$ values for transitions of  Ti XIX. ($a{\pm}b \equiv$ $a\times$10$^{{\pm}b}$).}      
\begin{tabular}{rrlllllllll}                                                                                    
\hline                                                                                                        
\hline 
\multicolumn{2}{c}{log T$_e$ (K)} &  \multicolumn{3}{c}{6.3} &   \multicolumn{3}{c}{6.9} &   
\multicolumn{3}{c}{7.5}  \\ 
\hline                                                                                                                                                                                                      
  $i$ & $j$    &   DARC       & FAC         &        ZS      &   DARC      &   FAC       &   ZS           &   DARC       &   FAC       &   ZS           \\					       
\hline                                          					      
    1  &    2  &   3.550$-$3  &  1.693$-$3  &	 1.7943$-$3  &  2.753$-$3  &  1.242$-$3  &    1.3135$-$3  &   1.240$-$3  &  6.553$-$4  &    6.9065$-$4  \\
    1  &    3  &   1.480$-$2  &  8.981$-$3  &	 9.1369$-$3  &  1.326$-$2  &  8.517$-$3  &    8.6113$-$3  &   9.330$-$3  &  8.208$-$3  &    8.1786$-$3  \\
    1  &    4  &   2.041$-$2  &  8.224$-$3  &	 8.7866$-$3  &  1.497$-$2  &  6.031$-$3  &    6.4270$-$3  &   6.568$-$3  &  3.180$-$3  &    3.3795$-$3  \\
    1  &    5  &   4.632$-$1  &  4.683$-$1  &	 4.4407$-$1  &  5.414$-$1  &  5.558$-$1  &    5.4721$-$1  &   6.535$-$1  &  7.149$-$1  &    7.1229$-$1  \\
    1  &    6  &   2.607$-$4  &  1.126$-$4  &	 1.1731$-$4  &  2.678$-$4  &  8.760$-$5  &    8.9602$-$5  &   1.421$-$4  &  6.029$-$5  &    5.9662$-$5  \\
    1  &    7  &   5.441$-$4  &  2.059$-$4  &	 2.2788$-$4  &  5.371$-$4  &  1.369$-$4  &    1.5205$-$4  &   2.272$-$4  &  6.298$-$5  &    6.7922$-$5  \\
    1  &    8  &   1.003$-$3  &  5.224$-$4  &	 4.8162$-$4  &  1.002$-$3  &  4.565$-$4  &    3.9854$-$4  &   6.263$-$4  &  4.007$-$4  &    3.2024$-$4  \\
    1  &    9  &   2.566$-$3  &  1.645$-$3  &	 1.2143$-$3  &  2.660$-$3  &  1.826$-$3  &    1.3591$-$3  &   2.379$-$3  &  2.116$-$3  &    1.5968$-$3  \\
    1  &   10  &   1.537$-$3  &  6.803$-$4  &	 5.7142$-$4  &  1.550$-$3  &  6.579$-$4  &    5.5271$-$4  &   1.102$-$3  &  6.290$-$4  &    5.3012$-$4  \\
    2  &    3  &   4.162$-$2  &  1.801$-$2  &	 1.9718$-$2  &  2.591$-$2  &  1.262$-$2  &    1.3622$-$2  &   1.118$-$2  &  6.353$-$3  &    6.8570$-$3  \\
    2  &    4  &   3.037$-$2  &  1.380$-$2  &	 1.3591$-$2  &  2.181$-$2  &  1.266$-$2  &    1.2352$-$2  &   1.490$-$2  &  1.192$-$2  &    1.1670$-$2  \\
    2  &    5  &   9.787$-$3  &  4.090$-$3  &	 4.4369$-$3  &  8.125$-$3  &  2.829$-$3  &    3.0501$-$3  &   3.445$-$3  &  1.387$-$3  &    1.4701$-$3  \\
    2  &    6  &   2.106$-$3  &  1.079$-$3  &	 1.1455$-$3  &  1.733$-$3  &  7.869$-$4  &    8.3760$-$4  &   7.891$-$4  &  4.119$-$4  &    4.3712$-$4  \\
    2  &    7  &   2.202$-$1  &  2.226$-$1  &	 2.2221$-$1  &  2.571$-$1  &  2.651$-$1  &    2.7186$-$1  &   3.058$-$1  &  3.409$-$1  &    3.4776$-$1  \\
    2  &    8  &   5.388$-$3  &  2.428$-$3  &	 2.5623$-$3  &  4.697$-$3  &  1.774$-$3  &    1.8632$-$3  &   2.101$-$3  &  9.312$-$4  &    9.7530$-$4  \\
    2  &    9  &   3.211$-$3  &  1.243$-$3  &	 1.3231$-$3  &  2.916$-$3  &  9.069$-$4  &    9.6379$-$4  &   1.262$-$3  &  4.755$-$4  &    5.0398$-$4  \\
    2  &   10  &   5.646$-$4  &  1.509$-$4  &	 1.5738$-$4  &  4.844$-$4  &  1.072$-$4  &    1.1212$-$4  &   1.976$-$4  &  5.397$-$5  &    5.6180$-$5  \\
    3  &    4  &   1.550$-$1  &  5.275$-$2  &	 5.4204$-$2  &  9.090$-$2  &  4.364$-$2  &    4.4300$-$2  &   4.970$-$2  &  3.436$-$2  &    3.4175$-$2  \\
    3  &    5  &   3.741$-$2  &  1.261$-$2  &	 1.3622$-$2  &  2.691$-$2  &  8.896$-$3  &    9.5108$-$3  &   1.147$-$2  &  4.685$-$3  &    4.9728$-$3  \\
    3  &    6  &   2.262$-$1  &  2.282$-$1  &	 2.2576$-$1  &  2.631$-$1  &  2.725$-$1  &    2.7867$-$1  &   3.112$-$1  &  3.505$-$1  &    3.5752$-$1  \\
    3  &    7  &   1.729$-$1  &  1.702$-$1  &	 1.6936$-$1  &  1.991$-$1  &  2.011$-$1  &    2.0605$-$1  &   2.307$-$1  &  2.564$-$1  &    2.6172$-$1  \\
    3  &    8  &   2.864$-$1  &  2.836$-$1  &	 2.8422$-$1  &  3.313$-$1  &  3.355$-$1  &    3.4509$-$1  &   3.849$-$1  &  4.288$-$1  &    4.3893$-$1  \\
    3  &    9  &   1.843$-$2  &  1.106$-$2  &	 1.0630$-$2  &  1.805$-$2  &  1.082$-$2  &    1.0263$-$2  &   1.354$-$2  &  1.111$-$2  &    1.0400$-$2  \\
    3  &   10  &   2.511$-$3  &  8.760$-$4  &	 9.0560$-$4  &  2.237$-$3  &  7.512$-$4  &    7.8159$-$4  &   1.196$-$3  &  6.309$-$4  &    6.5237$-$4  \\
    4  &    5  &   7.273$-$2  &  2.146$-$2  &	 2.3186$-$2  &  4.562$-$2  &  1.488$-$2  &    1.6003$-$2  &   1.844$-$2  &  7.398$-$3  &    7.8110$-$3  \\
    4  &    6  &   3.254$-$3  &  8.655$-$4  &	 9.4764$-$4  &  2.512$-$3  &  6.355$-$4  &    6.9369$-$4  &   1.052$-$3  &  3.358$-$4  &    3.6439$-$4  \\
    4  &    7  &   2.884$-$1  &  2.858$-$1  &	 2.8227$-$1  &  3.333$-$1  &  3.407$-$1  &    3.4819$-$1  &   3.890$-$1  &  4.371$-$1  &    4.4551$-$1  \\
    4  &    8  &   7.430$-$1  &  7.522$-$1  &	 7.5185$-$1  &  8.635$-$1  &  8.957$-$1  &    9.2247$-$1  &   1.014$+$0  &  1.150$+$0  &    1.1785$+$0  \\
    4  &    9  &   1.280$-$1  &  1.069$-$1  &	 1.0317$-$1  &  1.393$-$1  &  1.202$-$1  &    1.1693$-$1  &   1.469$-$1  &  1.465$-$1  &    1.4356$-$1  \\
    4  &   10  &   6.400$-$3  &  1.849$-$3  &	 1.9425$-$3  &  5.279$-$3  &  1.328$-$3  &    1.3970$-$3  &   2.164$-$3  &  6.808$-$4  &    7.1247$-$4  \\
    5  &    6  &   9.319$-$3  &  6.618$-$3  &	 7.2781$-$3  &  9.312$-$3  &  7.599$-$3  &    8.5691$-$3  &   8.387$-$3  &  9.108$-$3  &    1.0238$-$2  \\
    5  &    7  &   1.407$-$2  &  5.982$-$3  &	 6.2057$-$3  &  1.208$-$2  &  5.274$-$3  &    5.4767$-$3  &   6.677$-$3  &  4.463$-$3  &    4.6440$-$3  \\
    5  &    8  &   1.270$-$1  &  1.116$-$1  &	 1.1022$-$1  &  1.378$-$1  &  1.338$-$1  &    1.3584$-$1  &   1.462$-$1  &  1.687$-$1  &    1.7047$-$1  \\
    5  &    9  &   1.074$+$0  &  1.094$+$0  &	 1.0744$+$0  &  1.237$+$0  &  1.328$+$0  &    1.3820$+$0  &   1.422$+$0  &  1.717$+$0  &    1.7953$+$0  \\
    5  &   10  &   3.799$-$1  &  3.868$-$1  &	 3.9823$-$1  &  4.431$-$1  &  4.613$-$1  &    4.8814$-$1  &   5.268$-$1  &  5.939$-$1  &    6.2552$-$1  \\
    6  &    7  &   3.074$-$2  &  1.923$-$2  &	 2.0870$-$2  &  2.496$-$2  &  1.359$-$2  &    1.4672$-$2  &   1.144$-$2  &  6.908$-$3  &    7.3534$-$3  \\
    6  &    8  &   2.441$-$2  &  1.730$-$2  &	 1.7613$-$2  &  2.226$-$2  &  1.526$-$2  &    1.5500$-$2  &   1.620$-$2  &  1.347$-$2  &    1.3428$-$2  \\
    6  &    9  &   1.137$-$2  &  6.189$-$3  &	 6.9335$-$3  &  9.442$-$3  &  4.247$-$3  &    4.7452$-$3  &   4.080$-$3  &  2.062$-$3  &    2.2528$-$3  \\
    6  &   10  &   3.660$-$3  &  9.244$-$4  &	 1.0278$-$3  &  3.010$-$3  &  6.115$-$4  &    6.7632$-$4  &   1.169$-$3  &  2.795$-$4  &    3.0982$-$4  \\
    7  &    8  &   8.444$-$2  &  5.652$-$2  &	 5.9629$-$2  &  7.490$-$2  &  4.555$-$2  &    4.7635$-$2  &   4.542$-$2  &  3.381$-$2  &    3.4716$-$2  \\
    7  &    9  &   5.030$-$2  &  2.925$-$2  &	 3.1902$-$2  &  4.358$-$2  &  2.095$-$2  &    2.2555$-$2  &   2.079$-$2  &  1.163$-$2  &    1.2341$-$2  \\
    7  &   10  &   1.383$-$2  &  3.944$-$3  &		     &  1.088$-$2  &  2.632$-$3  &		  &   4.220$-$3  &  1.221$-$3  &		\\
    8  &    9  &   1.013$-$1  &  6.637$-$2  &	 7.0516$-$2  &  8.985$-$2  &  5.088$-$2  &    5.3039$-$2  &   5.028$-$2  &  3.388$-$2  &    3.3996$-$2  \\
    8  &   10  &   1.998$-$2  &  8.238$-$3  &	 8.7911$-$3  &  1.620$-$2  &  6.240$-$3  &    6.4961$-$3  &   7.948$-$3  &  4.246$-$3  &    4.2710$-$3  \\
    9  &   10  &   3.972$-$2  &  2.929$-$2  &	 2.8800$-$2  &  4.004$-$2  &  3.092$-$2  &    3.0604$-$2  &   3.769$-$2  &  3.457$-$2  &    3.4692$-$2  \\
\hline                                                                                                        
\end{tabular}  

\begin{flushleft}
{\small
DARC: Present calculations from the DARC code \\
FAC: Present calculations from the FAC code \\
ZS:   Calculations of Zhang and Sampson \cite{zs92} \\

}
\end{flushleft}
                                                                                               
\end{table*}                

\end{document}